\documentclass[acmsmall,screen]{acmart}

\AtBeginDocument{%
  }

\usepackage{amsmath,amsfonts}
\usepackage{algorithmic}
\usepackage{graphicx}
\usepackage{textcomp}
\usepackage{xcolor}
\usepackage{booktabs}
\usepackage{tikz,pgfplots}
\usepackage{standalone}
\usepackage[most]{tcolorbox}
\usepackage{csquotes}
\usepackage{url}
\usepackage{multirow}
\usetikzlibrary{arrows.meta}
\usepackage{fancyhdr}
\usepackage{threeparttable}
\usepackage{subcaption}
\usepackage{makecell}
\usepackage{hyperref}
\usepackage{newtxmath}
\usepackage{pgf-pie}
\usepackage{multirow}
\usepackage{fontawesome5}
\usepackage{pifont} 
\usepackage{xspace}

\definecolor{beaublue}{rgb}{0.74, 0.83, 0.9}
\definecolor{applegreen}{rgb}{0.55, 0.71, 0.0}
\definecolor{dollarbill}{rgb}{0.52, 0.73, 0.4}
\definecolor{ao(english)}{rgb}{0.0, 0.5, 0.0}
\definecolor{amaranth}{rgb}{0.9, 0.17, 0.31}
\definecolor{cerublue}{rgb}{0.16, 0.32, 0.75}
\definecolor{frenchblue}{rgb}{0.0, 0.45, 0.73}
\definecolor{iceberg}{rgb}{0.44, 0.65, 0.82}

\newcommand{\cmark}{\textcolor{applegreen}{\ding{51}}} 
\newcommand{\xmark}{\textcolor{amaranth}{\ding{55}}} 

\makeatletter
\tcbset{
    myhbox/.style 2 args={%
        enhanced, 
        breakable,
        colback=white,
        colframe=cerublue!70,
        attach boxed title to top left={yshift*=-\tcboxedtitleheight}, 
        title={#2},
        boxed title size=title,
        boxed title style={%
            sharp corners, 
            rounded corners=northwest, 
            colback=tcbcolframe, 
            boxrule=0pt,
        },
        underlay boxed title={%
            \path[fill=tcbcolframe] (title.south west)--(title.south east) 
                to[out=0, in=180] ([xshift=5mm]title.east)--
                (title.center-|frame.east)
                [rounded corners=\kvtcb@arc] |- 
                (frame.north) -| cycle; 
        },
        #1
    }
}   
\makeatother

\newtcolorbox{myhbox}[2][]{%
    myhbox={#1}{#2}
}

\newcounter{findingcount}
\newcounter{suggestioncount}
\newcommand{\keybox}[1]{
\begin{tcolorbox}[leftrule=1mm,toprule=0mm,bottomrule=0mm,left=1pt,right=2pt,top=2pt,bottom=2pt, colback=beaublue]
\em #1
\end{tcolorbox}
}

\newcommand{\suggestbox}[1]{
\begin{tcolorbox}[leftrule=1mm,toprule=0mm,bottomrule=0mm,left=1pt,right=2pt,top=2pt,bottom=2pt, colback=dollarbill!50]
\em #1
\end{tcolorbox}
}

\setcopyright{acmlicensed}
\copyrightyear{2024}
\acmYear{2024}
\acmDOI{XXXXXXX.XXXXXXX}

\acmJournal{CSUR}
\acmISBN{978-1-4503-XXXX-X/18/06}

\begin{document}

\title{Language Models for Code Optimization: Survey, Challenges and Future Directions}

\author{Jingzhi Gong}
\email{j.gong@leeds.ac.uk}
\orcid{0000-0003-4551-0701}
\affiliation{%
  \institution{University of Leeds}
  \city{Leeds}
  \country{UK}
}
\affiliation{%
  \institution{TurinTech AI}
  \city{London}
  \country{UK}
}

\author{Vardan Voskanyan}
\email{vardan@turintech.ai}
\orcid{0000-0002-3645-7710}
\affiliation{%
  \institution{TurinTech AI}
  \city{London}
  \country{UK}
}

\author{Paul Brookes}
\email{paul@turintech.ai}
\orcid{0000-0002-8639-8413}
\affiliation{%
  \institution{TurinTech AI}
  \city{London}
  \country{UK}
}

\author{Fan Wu}
\email{fan@turintech.ai}
\orcid{0000-0002-3734-7855}
\affiliation{%
  \institution{TurinTech AI}
  \city{London}
  \country{UK}
}

\author{Wei Jie}
\email{wei.jie@uwl.ac.uk}
\orcid{0000-0002-5392-0009}
\affiliation{%
  \institution{University of West London}
  \city{London}
  \country{UK}
}

\author{Jie Xu}
\email{j.xu@leeds.ac.uk}
\orcid{0000-0003-4102-233X}
\affiliation{%
  \institution{University of Leeds}
  \city{Leeds}
  \country{UK}
}


\author{Rafail Giavrimis}
\email{rafail@turintech.ai}
\orcid{0000-0002-5270-8065}
\affiliation{%
  \institution{University of Surrey}
  \city{Surrey}
  \country{UK}
}
\affiliation{%
  \institution{TurinTech AI}
  \city{London}
  \country{UK}
}

\author{Mike Basios}
\email{mike@turintech.ai}
\orcid{}
\affiliation{
  \institution{TurinTech AI}
  \city{London}
  \country{UK}
}

\author{Leslie Kanthan}
\email{leslie@turintech.ai}
\orcid{}
\affiliation{
  \institution{TurinTech AI}
  \city{London}
  \country{UK}
}

\author{Zheng Wang}
\email{z.wang5@leeds.ac.uk}
\authornote{Corresponding author}
\orcid{0000-0001-6157-0662}
\affiliation{%
  \institution{University of Leeds}
  \city{Leeds}
  \country{UK}
}
\renewcommand{\shortauthors}{Gong et al.}

\ccsdesc[500]{Software and its engineering~Software performance}

\keywords{Large Language Model, LLM, Code Performance Optimization, Code Optimisation, Code Performance Optimisation, Artificial Intelligence for Software Engineering, AI4SE}


\begin{abstract}
Language models (LMs) built upon deep neural networks (DNNs) have recently demonstrated breakthrough effectiveness in software engineering tasks like code generation, completion, and repair. This has paved the way for the emergence of LM-based code optimization techniques, which are crucial for enhancing the performance of existing programs, such as accelerating program execution time. However, a comprehensive survey dedicated to this specific application has been lacking. To fill this gap, we present a systematic literature review of over 50 primary studies, identifying emerging trends and addressing 11 specialized questions. Our findings reveal five critical open challenges, such as balancing model complexity with practical usability, cross-language/performance generalizability, and building trust in AI-driven solutions. Furthermore, we provide eight future research directions to facilitate more efficient, robust, and reliable LM-based code optimization. Thereby, this study aims to provide actionable insights and foundational references for both researchers and practitioners in this rapidly evolving field.
\end{abstract}

\maketitle

\section{Introduction}
\label{sec:introduction}
Code optimization, or program optimization, has long been an essential task in computing~\cite{wang2018machine}. Code optimization involves transforming a program at various levels---such as source code~\cite{DBLP:conf/iclr/ShypulaMZ0GYHNR24}, compiler intermediate representation~\cite{DBLP:journals/corr/abs-2407-02524}, or binary~\cite{fernandez1995simple,ben2020efficient,licker2020duplo}---to achieve specific performance goals like reducing execution time~\cite{DBLP:conf/nips/MadaanTGHGW0DPY23}, minimizing code size~\cite{rocha2019function,DBLP:conf/euromlsys/GrubisicSSLMC24}, or optimizing memory usage~\cite{DBLP:conf/sigsoft/GargMCSW22}. It underpins a wide range of software engineering (SE) tasks, including code generation~\cite{DBLP:conf/nips/Le0GSH22}, code repair~\cite{DBLP:conf/sigsoft/JinSTSLSS23}, code edits~\cite{DBLP:conf/sigsoft/GuptaKBCGKRS023}, and code refinement~\cite{DBLP:conf/acl/OpenCodeInterpreter24}. 

Traditionally, code optimization relied on expert-crafted heuristics and rules~\cite{wang2018machine}. These techniques were often integrated with compiler-based code analysis~\cite{DBLP:conf/pldi/YangCER11} to capture important program properties, such as data and control dependencies, to identify the most efficient ways to optimize the code. Over time, a wide range of optimization techniques has been developed, ranging from low-level strategies like instruction scheduling \cite{eigenmann2000parallelizing}, register allocation \cite{chaitin1982register}, vectorization~\cite{allen1988compiling}, and loop transformations~\cite{wolfe1982optimizing}---typically applied at the compiler's intermediate representation or during link-time optimization---to higher-level strategies for changing algorithms or data structures at the source code level for improved performance~\cite{DBLP:journals/nature/RomeraParedesBNBKDREWFKF24}.

One of the key challenges in code optimization is the vast number of possible ways to optimize an input program, making an exhaustive search computationally prohibitive, often taking many machine years to explore fully~\cite{DBLP:journals/nature/RomeraParedesBNBKDREWFKF24}. Within this vast space, good optimizations are usually sparse and can vary significantly between programs~\cite{DBLP:conf/euromlsys/GrubisicSSLMC24,wang2018machine}. For low-level performance optimization, the best optimizations are often dependent on the underlying computing hardware~\cite{voss2001high,cummins2017end}. This makes it highly challenging to manually craft an effective optimization strategy. Even if a well-tuned heuristic can be developed, it will likely require adaptation as both the application workload and the computing hardware evolve~\cite{curtis2006online}.


Over the past decades, a substantial body of work has explored the use of machine learning for code optimization~\cite{bhatsurvey, wang2018machine, ashouri2018survey}. There is now ample evidence showing the effectiveness of machine learning techniques across a wide range of code optimization tasks~\cite{wang2018machine}. More recently, the advent of language models (LMs) and generative artificial intelligence (GenAI), built on deep neural networks (DNNs), has marked a significant breakthrough in this area \cite{DBLP:conf/iclr/ShypulaMZ0GYHNR24}. These advanced models have demonstrated powerful capabilities in extracting knowledge from training data and transferring it to test samples \cite{DBLP:journals/corr/Gong24Deep}, outperforming classical machine learning approaches~\cite{cummins2017end}. Their ability to model and reason complex code structures has further spurred extensive research into leveraging LMs for software engineering~\cite{hou2024largelanguagemodelssoftware}, with promising results in automating and enhancing code optimization processes. This growing synergy between machine learning, LMs, and code optimization opens new avenues for research and innovation in the field.

Yet, despite the growing importance and promising advancements in code optimization using LMs, existing literature reviews on LMs in code-related tasks primarily focus on their general applications in software engineering~\cite{liu2024large} or specific domains like automatic program repair~\cite{zhang2024systematic}. Notably, there remains a significant gap in the literature---no comprehensive study has systematically reviewed LM-based techniques specifically for software code optimization.

\begin{figure}[!t]
\centering
\includegraphics[width=0.6\columnwidth]{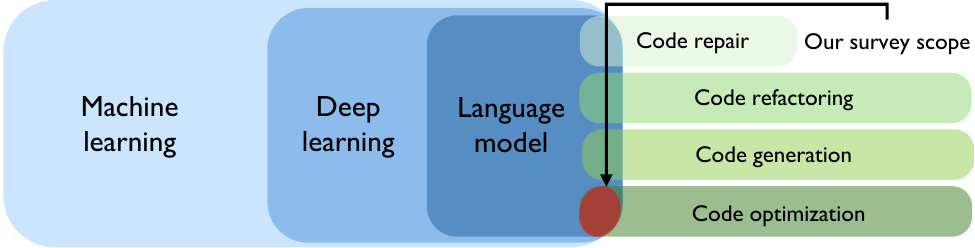}
    \caption{Visualization of the survey scope.}
    ~\vspace{-0.6cm}
 \label{fig:scope}
 \end{figure}

As depicted in Figure~\ref{fig:scope}, this paper aims to fill this gap by offering a systematic literature review (SLR) of state-of-the-art LM-based approaches for code optimization. Specifically, by searching through six academic indexing engines, we identified and systematically reviewed 53 primary studies\footnote{Full list of studies and all the raw results of this survey can be accessed at: \textcolor{blue}{\url{https://github.com/gjz78910/CodeOpt-SLR}}.}. Based on four research questions (RQs) with 11 specified sub-questions, we categorized these studies, summarized key findings, and offered insightful recommendations for readers. For example, our main findings include: 
\begin{itemize}
    \item General-purpose LMs like GPT-4 were more widely adopted (61 instances) than code-specialized LMs (43 instances) due to their broader understanding and reasoning capabilities.
    \item A majority of studies (57\%) leveraged pre-trained models to save time and resources, while 43\% employed fine-tuning to tailor the models for task-specific needs.
    \item The most commonly highlighted challenges were performance and code-related, such as limitation of one-step optimization (18 studies), balancing correctness and efficiency (15 studies), and complexity of code syntax (10 studies).
    \item Most studies addressed existing challenges by building dedicated models (51 instances), which are effective but lack generalizability. Prompt engineering stood out as the second category (34 instances) for its data efficiency, albeit reliant on expert knowledge. Another category formulated new problems for code optimization (33 instances), offering greater flexibility but demanding extensive effort in dataset preparation. 
\end{itemize}

Furthermore, we revealed five key challenges in the existing literature and provided potential directions for future research, in summary:
\begin{itemize}
    \item The increasing size and complexity of LMs demand significant computational resources for optimizing large-scale codebases, posing requirements for model compression and ensembling techniques.
    \item LM-based code optimization methods often operate in isolated environments, lacking seamless integration with external systems, underscoring the importance of agentic LMs.
    \item  The dominance of single-language studies (81\%) and the emphasis on single performance metrics (79\%) highlight the challenges in generalizability and the need for multi-lingual and multi-objective optimization approaches.
    \item Most LM-based methods were evaluated on synthetic datasets (68\%) rather than real-world codebases that are typically larger and more complex, indicating the necessity of standardized benchmarks reflecting different real-world scenarios.
    \item LMs often produce inconsistent or hallucinated outputs, making human-LM collaboration essential to harness AI's computational power while ensuring trustworthiness and reliability in optimization results.
\end{itemize}

The rest of the paper is organized as follows: Section~\ref{sec:background} illustrates the evolution of code optimization techniques. Section~\ref{sec:methodology} outlines the SLR methodology employed. Sections~\ref{sec:RQ1_characteristics_LMs}, \ref{sec:RQ2_LM_application}, \ref{sec:RQ3_problem_definition}, and \ref{sec:RQ4_evaluation_method} present the results and findings of the four research questions. Section~\ref{sec:challenges_future_directions} discusses existing challenges and future directions. 
Finally, Section~\ref{sec:conclusion} concludes the paper.

\section{Background}
\label{sec:background}
This section outlines the related concepts and development history of code optimization methods.

\subsection{Code Optimization}
\begin{figure}[t!]
\centering
\begin{subfigure}{0.31\textwidth}
\begin{lstlisting}
total = 0
for i in range(1, n+1):
    total += i
# Time complexity of O(n)
\end{lstlisting}
\caption{Unoptimized Python code}
\label{subfig:unoptimized_code}
\end{subfigure}
\hspace{1.5cm}
\begin{subfigure}{0.31\textwidth}
\begin{lstlisting}
total = n * (n + 1) // 2 
# Time complexity of O(1)
\end{lstlisting}
\caption{An optimized version of \ref{subfig:unoptimized_code}}
\end{subfigure}
~\vspace{-0.1cm}
\caption{Two Python implementations for calculating the sum of the first \( n \) natural numbers.}
~\vspace{-0.6cm}
\label{fig:code_opt_example}
\end{figure}

This article focuses on code optimization techniques that enhance performance objectives while preserving the original functionality, such as achieving faster execution speed or reducing binary code size and memory usage. Optimization can be applied at multiple levels, including the source code, intermediate representation (IR), and binary levels.
At the source code level, changes to algorithms, data structures, or implementation details can significantly improve performance~\cite{DBLP:conf/iclr/ShypulaMZ0GYHNR24}. For instance, Figure~\ref{fig:code_opt_example} demonstrates how an unoptimized Python program (Figure~\ref{fig:code_opt_example}a) can be optimized by replacing a loop-based summation with a direct computation (Figure~\ref{fig:code_opt_example}b). A range of optimization and analysis techniques like dead code elimination, loop unrolling, and vectorization can be applied at the IR level to reduce redundant computation and exploit hardware features~\cite{voss2001high}. At the binary level, link-time optimizations such as instruction scheduling and memory layout optimization further improve computation and memory access efficiencies~\cite{fernandez1995simple}.

One of the key challenges in code optimization lies in navigating the vast optimization space, which contains numerous potential code transformation options \cite{park2013predictive}. Good and bad solutions exist within this space, often depending on the specific input program and the underlying hardware \cite{li2021analytical}. Effective code optimization methods must have robust strategies to explore this complex space and identify high-performing solutions. Traditionally, this was achieved through expert-crafted heuristics \cite{wang2018machine}, analytical models \cite{cavazos2006automatic}, or conventional machine learning approaches.
With the recent breakthroughs of DNNs and LMs in navigating complex decision spaces,
there is a growing interest in leveraging LMs for code optimization tasks, due to their superior generalization capabilities and the ability to generate human-readable explanations \cite{DBLP:conf/iclr/ShypulaMZ0GYHNR24}. 

While code optimization primarily aims to enhance the performance of software, several related activities play a vital role in the broader software development process\footnote{Due to space limitation, the full related works section can be accessed at: \textcolor{blue}{\url{https://github.com/gjz78910/CodeOpt-SLR}}.}. These include code generation, which refers to the automatic production of code from structured specifications or natural language descriptions, often without relying on an existing codebase \cite{DBLP:journals/corr/abs-2203-07814}; code refactoring, which focuses on improving the internal structure and readability of the code without altering its external behavior or performance \cite{li2023codeeditor}; and code repair, which involves modifying code to fix bugs or introduce new features, ensuring its correct functionality \cite{DBLP:conf/sigsoft/JinSTSLSS23}. 

In this survey, we focus on code optimization due to its direct impact on performance metrics that are critical in numerous applications, such as reducing execution time and memory usage. Moreover, optimized code facilitates the development of faster and more efficient software---a key consideration in resource-constrained environments and a significant competitive advantage in fields like high-performance computing \cite{palkowski2024gpt}.

\subsection{Code Optimization Methods Development History}
\begin{figure}[!t]
\centering
\includegraphics[width=0.9\columnwidth]{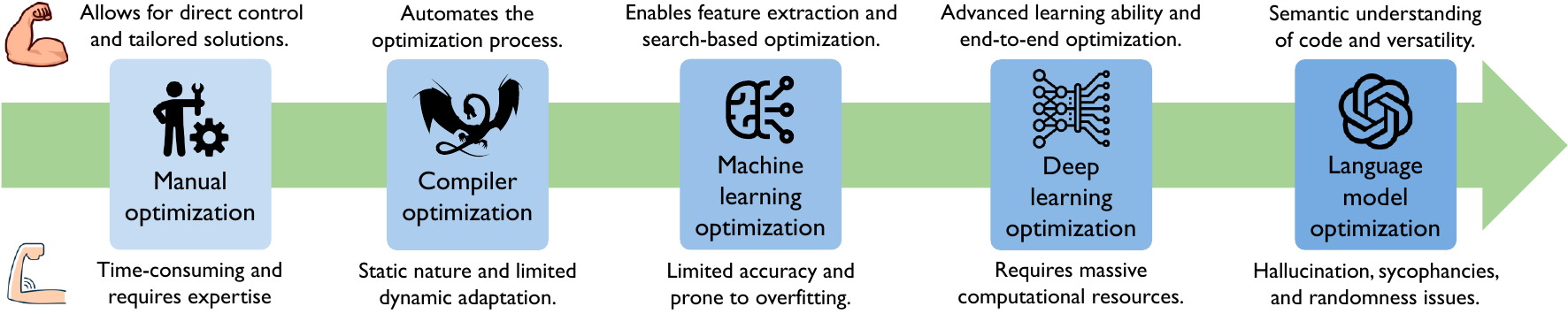}
~\vspace{-0.2cm}
    \caption{Development of code optimization methods: strengths and weaknesses}
 \label{fig:history}
 \end{figure}

Code optimization has been a key aspect of software development since the early days of computing~\cite{lowry1969object}. Figure~\ref{fig:history} traces its evolution, showcasing key methods, their strengths, and limitations. Early approaches centered on manual optimizations~\cite{curtis2006online}, where developers used assembly code to write and optimize software. Techniques like loop unrolling, function inlining, and minimizing memory access were developed to enhance speed and reduce memory usage~\cite{curtis2006online}.

The rise of high-level programming languages shifted optimization responsibilities to compilers~\cite{backus1978history}. Modern compilers integrate a wide range of code optimization techniques. These include instruction-level optimizations like peephole optimization~\cite{mckeeman1965peephole}, which refines small instruction sequences into more efficient forms. Moreover, loop optimizations \cite{allen1988compiling}, including loop unrolling and loop fusion, reduced loop control overhead, while inlining \cite{cammarota2013determination} replaced function calls with the function's body, cutting down call overhead.

In the last few decades, machine learning (ML) has been increasingly employed for code optimization, taking a data-driven approach to improve the efficiency of code. 
Classical ML techniques use carefully crafted feature extraction methods to capture code characteristics to identify performance bottlenecks and guide optimization decisions \cite{sikka2020learning}. Additionally, ML models can predict the performance of alternative code paths, serving as utility functions to navigate the optimization space and identify transformations that meet performance goals \cite{baghdadi2021deep}. In particular, \citet{DBLP:conf/scam/AdlerFGKLLLS21} proposed a search-based approach to enhance the readability of SCRATCH programs. ML-based code optimizations have also been adopted by the open-source community~\cite{fursin2011milepost, chen2018tvm, trofin2021mlgo} and industry. For example, Artemis++ \cite{DBLP:conf/kbse/GiavrimisBPB021} employs mutation algorithms to generate optimized C++ code, improving runtime, CPU, and memory usage. 

Deep learning (DL), a subset of ML, further advances code optimization by leveraging neural networks to model complex code relationships. DL models automatically learn code representations that capture semantic meaning, revealing optimization opportunities beyond traditional analysis \cite{baghdadi2021deep}. End-to-end approaches have gained traction, where DL models optimize code from source to executable. For example, DeepTune \cite{cummins2017end} uses deep neural networks to build optimization heuristics directly from raw source code, bypassing manual feature extraction. This approach allows DL models to learn from large code and performance metrics datasets, streamlining the end-to-end optimization process.

These methods, while effective in certain scenarios, do not always meet the specific needs of diverse use cases due to their lack of flexibility and the complexity involved in understanding and debugging automated optimizations. Specifically, manual optimization requires deep knowledge of both hardware and software, and compiler optimizations might not always produce the most efficient code, and their static nature prevents adaptation to runtime conditions~\cite{DBLP:journals/corr/abs-2309-14846}. Machine and deep learning models, although more advanced, rely heavily on the quality of feature extraction and training data, limiting their ability to capture the complex semantics and contextual relationships within code~\cite{bhatsurvey}. Additionally, search-based algorithms can be time-consuming and may not always converge to the optimal solution~\cite{DBLP:journals/nature/RomeraParedesBNBKDREWFKF24}. 

\subsection{Code Optimization Using LMs}
The recent advent of LMs has brought a paradigm shift in code optimization methods to LMs. The advantages of using LMs for code optimization are numerous. Firstly, a key advantage of LMs is their deep semantic understanding of code. By training on extensive datasets comprising code, functionality, comments, and documentation, LMs acquire the ability to reason about program logic. This capability allows them to outperform traditional machine learning models, enabling complex tasks such as loop restructuring, elimination of redundant computations, and memory access optimization---all aligned with user objectives. For instance,~\citet{DBLP:journals/corr/abs-2203-07814} introduced AlphaCode, a model capable of generating diverse programs, filtering, and classifying them to identify optimal solutions. This approach demonstrated human-level performance in solving competitive programming problems. Similarly, LM-based tools like GitHub Copilot~\cite{openai2021codex}, powered by OpenAI’s Codex, offer practical support for code optimization by providing real-time suggestions and auto-completions within Integrated Development Environments (IDEs).

Another major strength of LMs is their capacity to explore the optimization space. Unlike manual or compiler-based approaches, which depend on static rules, predefined heuristics, or handcrafted features, LMs exhibit greater adaptability and flexibility. By training on large-scale datasets that encompass a broad range of programming languages, paradigms, and performance scenarios, LMs dynamically reason, generate, and optimize code, identifying optimization opportunities that static methods often overlook~\cite{DBLP:journals/nature/RomeraParedesBNBKDREWFKF24}. For instance,~\citet{DBLP:conf/gi-ws/KangY23} demonstrated how LMs could enhance inefficient implementations of Fibonacci sequence calculations, showcasing their role as mutators in Genetic Improvement (GI) to produce mutants tailored to specific objectives.

Furthermore, LMs offer remarkable versatility in supporting various code optimization tasks. They can directly generate optimized code for a target language as an \textit{optimizer}~\cite{DBLP:journals/corr/abs-2401-14196}, provide natural language or mixed code-natural language suggestions as an \textit{advisor}~\cite{sun2024autosat}, or serve as an \textit{encoder} to transform code into feature vectors for downstream machine learning models~\cite{DBLP:conf/emnlp/FengGTDFGS0LJZ20}. Additionally, LMs can act as \textit{evaluators}, predicting the potential benefits of specific code transformations to guide optimization strategies~\cite{hemberg2024evolving}. This multi-faceted functionality makes LMs a vital tool for modern code optimization.

\begin{figure}[!t]
\centering
\includegraphics[width=\columnwidth]{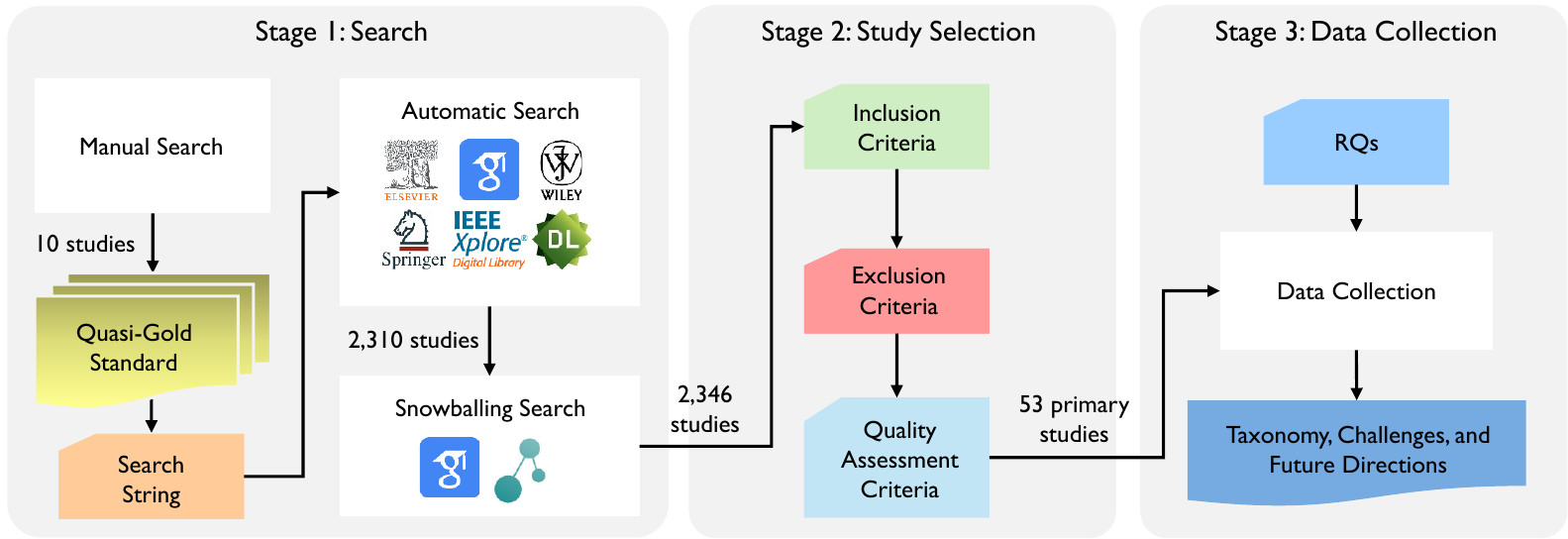}
    \caption{Overview of the survey methodology used in this study.}
 \label{fig:methodology}
 \end{figure}
 



However, there are limitations to using LMs for code optimization. One major challenge is the computational resources required to train and run these models, which can be substantial~\cite{minaee2024large}. The effectiveness of LMs also depends on the quality and diversity of the training data they have been exposed to~\cite{hou2024largelanguagemodelssoftware}. Further, integrating LMs into the development workflow can add complexity and require specialized effort, which might be a barrier for some small teams~\cite{DBLP:conf/fose-ws/FanGHLSYZ23}. Additionally, while LMs are powerful, they are not infallible and can sometimes suggest suboptimal or incorrect optimizations, necessitating human oversight~\cite{M2024humanAI}. This limitation is further highlighted by an empirical study on the optimization capabilities of LMs against traditional optimizing compilers, which discovered that even though LMs show large potential in code optimization, they currently struggle with larger programs and often yield marginal improvements over traditional compilers or code optimization tools \cite{DBLP:journals/corr/abs-2406-12146}.

Therefore, to understand and harness the full potential of these advanced LMs, our aim is to conduct a comprehensive survey at the intersection of LMs and code optimization, as illustrated in Figure~\ref{fig:scope}. By systematically studying the capabilities and limitations of LMs for code optimization, researchers and practitioners can develop more effective strategies for integrating these models into the software development lifecycle, improve software performance, resource efficiency and developer productivity, ultimately advancing the field of software engineering.



\section{Methodology}
\label{sec:methodology}

This survey follows the widely recognized guidelines for SLRs in Software Engineering proposed by~\citet{keele2007guidelines}, which have also been adopted by numerous SLRs~\cite{DBLP:journals/corr/Gong24Deep, hou2024largelanguagemodelssoftware, DBLP:journals/tse/WangHGGZFSLZN23, zhang2024systematic, DBLP:journals/corr/abs-2405-04760}. As shown in Figure~\ref{fig:methodology}, the methodology encompasses three key stages\footnote{Due to space limitation, the full methodology can be accessed in our repository: \textcolor{blue}{\url{https://github.com/gjz78910/CodeOpt-SLR}}.}: 
\begin{enumerate}
    \item \emph{Search}: comprehensive automatic searches were conducted, using a carefully defined search string following the ``quasi-gold standard'' methodology~\cite{DBLP:journals/infsof/ZhangBT11}, supplemented by snowballing searches to ensure broad coverage.
    \item \emph{Study selection}: the searched studies were filtered using rigorous inclusion and exclusion criteria, followed by quality assessments to include only reliable and high-quality studies.
    \item \emph{Data collection}: four main RQs, comprising 11 specialized questions, were formulated to guide data extraction and analysis, leading to the primary outcomes of the survey.
\end{enumerate}

\begin{figure}[!t]
\centering
\includegraphics[width=0.85\columnwidth]{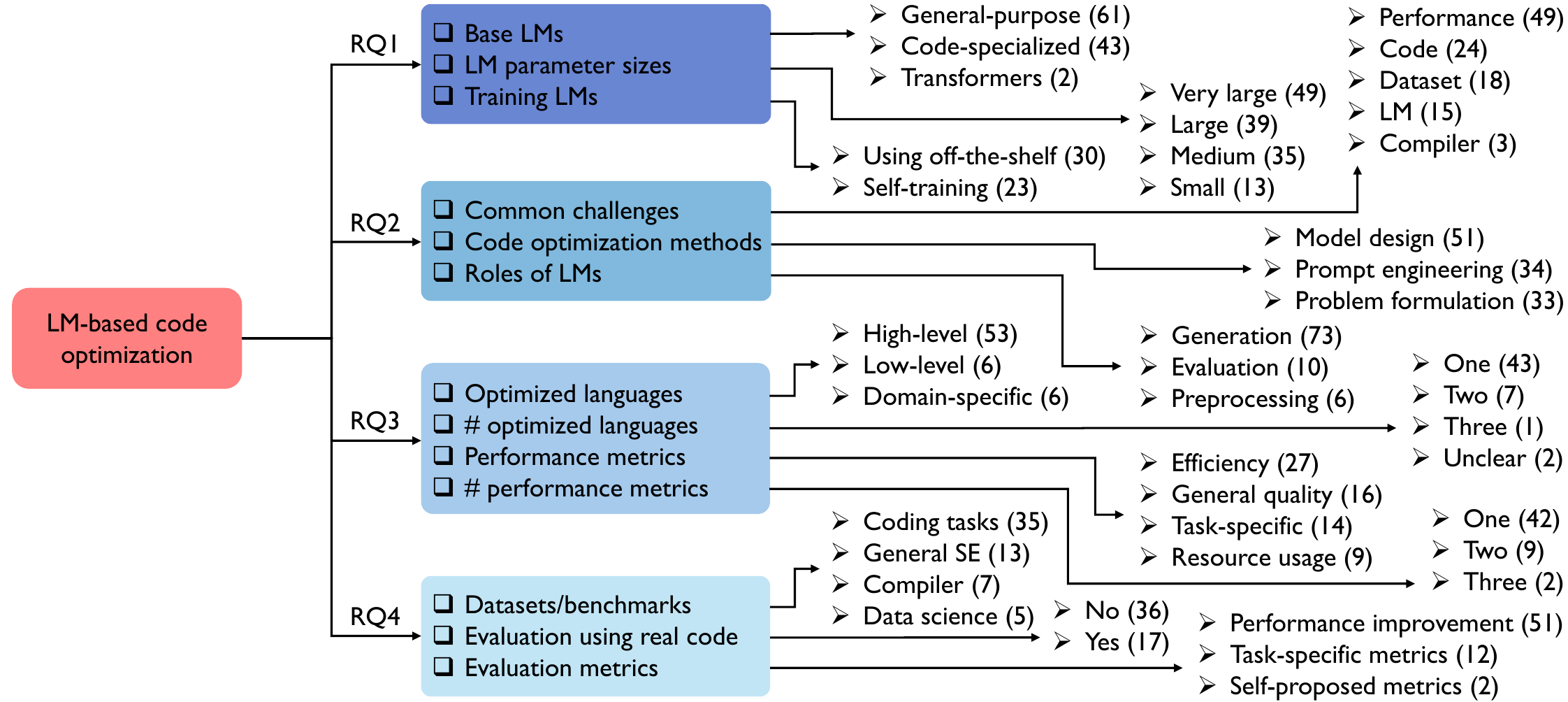}
    \caption{Overview of the taxonomy for all RQs (one study might be in multiple categories).}
    ~\vspace{-0.6cm}
 \label{fig:overview}
 \end{figure}

Figure~\ref{fig:overview} provides an overview of the taxonomy for all questions, and in the following sections, we will introduce the detailed taxonomy, findings, and actionable suggestions for each RQ separately.

\section{RQ1: What Were the Characteristics of the LMs Used for Code Optimization?}
\label{sec:RQ1_characteristics_LMs}
Despite the extensive use of LMs for code optimization tasks, there remains a notable gap in understanding how these models can be leveraged effectively. In this section, we investigate a few characteristics of the LMs used, introduce a detailed taxonomy for each sub-RQ, and discuss their implications in code optimization.

\begin{table}[t!]
  \caption{Distribution of LMs used for code optimization (one study might be in multiple categories).}
\centering
\footnotesize
\begin{adjustbox}{width=\linewidth,center}
\begin{tabular}{p{1cm}p{0.5cm}p{2.8cm}p{1.5cm}p{0.8cm}p{0.8cm}p{9cm}p{0.3cm}p{4cm}}
\toprule
\textbf{Category} & \textbf{Total \#} & \textbf{LM} & \textbf{Parameter size} & \textbf{Open source} & \textbf{Release year} & \textbf{Description} & \textbf{\#} & \textbf{Used studies} \\ \hline
\multirow{16}{1cm}{General-purpose LMs} & \multirow{16}{*}{61} & GPT-4~\cite{openai2024gpt4technicalreport} & $\approx$1.8T & \xmark & 2024 & The vast parameter size and   extensive training data enables its improved reasoning abilities and the   ability to process more complex instructions. & 15 & \cite{DBLP:journals/corr/abs-2310-02304,   gao2024searchbasedllmscodeoptimization,   shivashankar2024betterpythonprogrammingall,   DBLP:journals/corr/abs-2405-15189, DBLP:conf/nips/MadaanTGHGW0DPY23,   DBLP:conf/nips/ShinnCGNY23, DBLP:journals/corr/abs-2401-08500,   DBLP:journals/corr/abs-2406-04693, DBLP:conf/iclr/ShypulaMZ0GYHNR24,   DBLP:journals/concurrency/HanLDLZ24, DBLP:journals/pacmpl/YeZSS23,   xu2024large, sun2024autosat, zhang2024revolve, peng2024perfcodegen} \\ \cline{3-9} 
 &  & GPT-3.5-turbo~\cite{openai2022chatgpt} & $\approx$175B & \xmark & 2022 & Faster response times and more   cost-efficient compared to GPT-3.5. & 9 & \cite{DBLP:journals/corr/abs-2405-15189,   DBLP:conf/nips/MadaanTGHGW0DPY23, DBLP:conf/apsec/ShirafujiOSMW23,   DBLP:journals/corr/abs-2306-17077,   ishida2024langpropcodeoptimizationframework, xu2024large, van2024llamea,   hemberg2024evolving, choi2024iterative} \\ \cline{3-9} 
 &  & GPT   3.5~\cite{openai2022chatgpt} & $\approx$175B & \xmark & 2022 & An earlier version of GPT-4,   known for its solid capability in understanding and generating human-like   text and code. & 9 & \cite{pan2024measuringcodeefficiencyoptimization,   shivashankar2024betterpythonprogrammingall, DBLP:conf/nips/MadaanTGHGW0DPY23,   DBLP:journals/corr/abs-2401-08500, DBLP:conf/iclr/ShypulaMZ0GYHNR24,   gao2024searchbasedllmscodeoptimization, xu2024optimizing, palkowski2024gpt,   peng2024perfcodegen} \\ \cline{3-9} 
 &  & GPT-4o~\cite{openai2024gpt4o} & $\approx$1.8T & \xmark & 2024 & A multi-modal version of   GPT-4  that can handle multimodal code   contexts. & 7 & \cite{DBLP:journals/corr/abs-2406-07496,   DBLP:journals/corr/abs-2407-14044, peng2024large, wei2024improving,   van2024llamea, zhang2024revolve, van2024loop} \\ \cline{3-9} 
 &  & GPT-4-turbo~\cite{openai2024gpt4technicalreport} & $\approx$1.8T & \xmark & 2024 & Combines the strengths of GPT-4   with improved efficiency for faster processing. & 4 & \cite{DBLP:journals/corr/abs-2408-03408,   DBLP:journals/corr/abs-2405-15189,   yao2024rtlrewritermethodologieslargemodels, van2024llamea} \\ \cline{3-9} 
 &  & LLaMA-2~\cite{DBLP:journals/corr/llama2} & 7B, 13B, 34B & \cmark & 2023 & Enhanced capabilities and   efficiency over LLaMA-1. & 4 & \cite{li2024instructcoderinstructiontuninglarge,   DBLP:conf/euromlsys/GrubisicSSLMC24, DBLP:journals/corr/abs-2309-07062,   DBLP:journals/corr/abs-2403-14714} \\ \cline{3-9} 
 &  & Claude-3-haiku~\cite{claude2024} & $\approx$20B & \xmark & 2024 & Fastest among the Claude-3   models, optimized for near-instant responsiveness. & 2 & \cite{DBLP:journals/corr/abs-2405-15189,   han2024generating} \\ \cline{3-9} 
 &  & Gemini-Pro~\cite{gemini2023} & $\approx$540B & \xmark & 2023 & Google's multimodal model, like   GPT-4o, leveraging the MoE architecture. & 2 & \cite{DBLP:journals/corr/abs-2404-18864,   gao2024searchbasedllmscodeoptimization} \\ \cline{3-9} 
 &  & LLaMA-3.1~\cite{meta2024llama3} & 8B & \cmark & 2024 & Improves over LLaMA-2 with   expanded context length and multilingual support. & 2 & \cite{zhang2024revolve,   peng2024perfcodegen} \\ \cline{3-9} 
 &  & Claude-3-sonnet~\cite{claude2024} & $\approx$70B & \xmark & 2024 & Larger than Claude-3-haiku,   providing stronger performance and precision. & 1 & \cite{DBLP:journals/corr/abs-2405-15189} \\ \cline{3-9} 
 &  & LLaMA-1~\cite{DBLP:journals/corr/llama1} & 7B, 13B, 34B & \cmark & 2023 & An open-source LM that can be   fine-tuned for code optimization. & 1 & \cite{li2024instructcoderinstructiontuninglarge} \\ \cline{3-9} 
 &  & PaLM-2~\cite{anil2023palm} & 340B & \xmark & 2023 & Excels at solving complex tasks   by  decomposing them into simpler   subtasks. & 1 & \cite{xu2024large} \\ \cline{3-9} 
 &  & Phi-2~\cite{javaheripi2023phi} & 2.7B & \cmark & 2023 & Achieves remarkable performance   despite its relatively compact size. & 1 & \cite{zhang2024codedpo} \\ \cline{3-9} 
 &  & BLOOM~\cite{DBLP:journals/corr/bloom2022} & 3B, 7B & \cmark & 2022 & A multilingual language model   designed for general text processing. & 1 & \cite{li2024instructcoderinstructiontuninglarge} \\ \cline{3-9} 
 &  & GPT-NeoX~\cite{black2022gptneox20bopensourceautoregressivelanguage} & 20B & \cmark & 2022 & Provides accurate and   contextually relevant responses for text processing tasks. & 1 & \cite{pan2024ecodemasteringefficientcode} \\ \cline{3-9} 
 &  & GPT-3~\cite{brown2020language} & $\approx$175B & \xmark & 2020 & Ealier version of GPT-3.5, known   for its general NLP abilities. & 1 & \cite{jain2022jigsaw} \\ \hline
\multirow{17}{1cm}{Code-specialized LMs} & \multirow{17}{*}{43} & Code LLaMA~\cite{codellama2023} & 7B, 13B, 34B, 70B & \cmark & 2023 & A LLaMA model fine-tuned for   strong code-related performance, benefiting from the efficiency and   architecture of LLaMA. & 11 & \cite{li2024instructcoderinstructiontuninglarge,   DBLP:conf/iclr/ShypulaMZ0GYHNR24, DBLP:journals/corr/abs-2406-11935,   DBLP:journals/corr/abs-2407-14044, DBLP:journals/corr/abs-2406-12502,   DBLP:journals/corr/abs-2407-02524, gao2024searchbasedllmscodeoptimization,   DBLP:journals/corr/abs-2405-15189, xu2024optimizing, qu2024dynamic,   xu2024code} \\ \cline{3-9} 
 &  & DeepSeekCoder~\cite{deepseekcoder2023} & 1.3B, 6.7B, 33B & \cmark & 2023 & Shows competitive performance in   coding tasks due to its incorporation of semantic search and retrieval   mechanisms. & 7 & \cite{DBLP:journals/corr/abs-2405-15189,   DBLP:journals/corr/abs-2401-08500, DBLP:journals/corr/abs-2404-18864,   DBLP:journals/corr/abs-2406-11935, DBLP:journals/corr/abs-2407-14044,   zhang2024codedpo, huang2024effi} \\ \cline{3-9} 
 &  & StarCoder~\cite{DBLP:journals/tmlr/LiAZMKMMALCLZZW23} & 1B, 3B, 7B, 15B & \cmark & 2023 & Trained on a massive dataset of   permissively licensed source code, making it more readily usable in   commercial applications. & 4 & \cite{DBLP:journals/corr/abs-2405-15189,   DBLP:journals/nature/RomeraParedesBNBKDREWFKF24,   DBLP:journals/corr/abs-2407-14044, DBLP:journals/corr/abs-2406-12502} \\ \cline{3-9} 
 &  & CodeT5~\cite{DBLP:conf/emnlp/2021codeT5} & 60M, 220M, 770M & \cmark & 2021 & T5 model fine-tuned for coding   tasks, offering a balance of general language understanding and code   specialization. & 4 & \cite{pan2024measuringcodeefficiencyoptimization,   DBLP:journals/corr/abs-2312-05657, DBLP:journals/pacmpl/YeZSS23,   li2024falcon} \\ \cline{3-9} 
 &  & WizardCoder~\cite{DBLP:conf/iclr/wizardcoder24} & 13B & \cmark & 2024 & Improved coding capabilities due   to the Evol-Instruct training method. & 3 & \cite{shivashankar2024betterpythonprogrammingall,   DBLP:journals/corr/abs-2405-15189, DBLP:journals/corr/abs-2407-14044} \\ \cline{3-9} 
 &  & Qwen2.5-Code~\cite{hui2024qwen2} & 7B & \cmark & 2024 & Provides advanced coding   assistance and improves productivity for developers. & 2 & \cite{huang2024effi, xu2024code} \\ \cline{3-9} 
 &  & CodeX~\cite{openai2021codex} & 12B & \cmark & 2021 & A powerful coding assistant that   is integrated with GitHub Copilot. & 2 & \cite{DBLP:conf/nips/MadaanTGHGW0DPY23,   jain2022jigsaw} \\ \cline{3-9} 
 &  & StarCoder2~\cite{lozhkov2024starcoder} & 7B & \cmark & 2024 & Trained on significantly larger   and more diverse coding data than StarCoder. & 1 & \cite{zhang2024codedpo} \\ \cline{3-9} 
 &  & CodeGemma~\cite{DBLP:journals/corr/codeGemma2024} & 7B & \cmark & 2024 & Optimized for coding tasks using   pre-trained Gemma models. & 1 & \cite{DBLP:journals/corr/abs-2407-14044} \\ \cline{3-9} 
 &  & OpenCodeInterpreter~\cite{DBLP:conf/acl/OpenCodeInterpreter24} & 1.3B, 6.7B, 33B & \cmark & 2024 & Combines a language model with a   code execution environment, allowing it to optimize code by directly   evaluating its performance. & 1 & \cite{DBLP:journals/corr/abs-2405-15189} \\ \cline{3-9} 
 &  & Codey~\cite{google2023codey} & 340B & \cmark & 2023 & Provides code suggestions,   completions, and refactoring assistance. & 1 & \cite{DBLP:journals/nature/RomeraParedesBNBKDREWFKF24} \\ \cline{3-9} 
 &  & XwinCoder~\cite{DBLP:journals/corr/xwincoder} & 7B, 13B, 34B & \cmark & 2023 & Focuses on cross-lingual code   understanding and generation. & 1 & \cite{DBLP:journals/corr/abs-2405-15189} \\ \cline{3-9} 
 &  & CodeGen-mono~\cite{DBLP:conf/iclr/codegen23} & 350M & \cmark & 2023 & Achieves superior coding   accuracy by focusing exclusively on one language & 1 & \cite{pan2024measuringcodeefficiencyoptimization} \\ \cline{3-9} 
 &  & PolyCoder~\cite{DBLP:conf/pldi/polycoder22} & 400M & \cmark & 2022 & Emphasizing multilingual   programming capabilities & 1 & \cite{pan2024measuringcodeefficiencyoptimization} \\ \cline{3-9} 
 &  & CodeBERT~\cite{DBLP:conf/emnlp/FengGTDFGS0LJZ20} & 125M & \cmark & 2020 & Leverages BERT architecture for   better understanding of code semantics. & 1 & \cite{DBLP:journals/corr/abs-2309-14846} \\ \cline{3-9} 
 &  & PyMT5~\cite{DBLP:conf/emnlp/2020pymt5} & 374M & \cmark & 2020 & Optimized for Python code,   providing targeted code improvements. & 1 & \cite{DBLP:conf/sigsoft/GargMCSW22} \\ \cline{3-9} 
 &  & TransCoder~\cite{DBLP:conf/nips/RoziereLCL20} & $\approx$60M & \cmark & 2020 & Specialized in translating code   between programming languages. & 1 & \cite{DBLP:conf/hpec/GuoM22} \\ \hline
\multirow{2}{1cm}{Trans-formers} & \multirow{2}{*}{2} & Bert-tiny~\cite{turc2019wellreadstudentslearnbetter} & 4.4M & \cmark & 2019 & A smaller version of BERT,   suitable for scenarios requiring fast response times. & 1 & \cite{pan2024ecodemasteringefficientcode} \\ \cline{3-9} 
 &  & Transformer~\cite{DBLP:conf/nips/VaswaniSPUJGKP17} & $\approx$30M & \cmark & 2017 & The foundational architecture   for many LMs. & 1 & \cite{DBLP:journals/corr/abs-2109-13498}
\\
\bottomrule
\end{tabular}
\end{adjustbox}
\label{tb:base_lms}
\end{table}

\subsection{RQ1.1: Which LMs Were Used?}
\label{subsec:base_LMs}
Unlike surveys specializing in LMs~\cite{zhao2023survey, patil2024review, minaee2024large} that aim to provide a comprehensive list of LM architectures, our target in this subsection is to illustrate the characteristics of the foundation LMs used for code optimization, providing guidelines and insights for future researchers to select their most suitable LMs. As shown in Table~\ref{tb:base_lms}, they can be categorized into three main classes. 

\subsubsection{General-purpose LMs} 
A total of 61 general-purpose LMs were utilized in the primary studies, which are designed for a variety of tasks beyond code optimization. Among these, the most popular were various versions of GPT-4 and GPT-3.5---two successive versions of OpenAI's generative LMs---with GPT-4 being the most widely used, appearing in 15 studies~\cite{shivashankar2024betterpythonprogrammingall,   DBLP:journals/corr/abs-2405-15189, DBLP:conf/nips/MadaanTGHGW0DPY23,   DBLP:conf/nips/ShinnCGNY23}. As noted by~\citet{DBLP:journals/corr/abs-2406-04693}, GPT-4 demonstrated superior capabilities in contextual understanding and reasoning, making it particularly effective in applications requiring advanced code comprehension and optimization. Besides, studies also employed other general-purpose LMs from the GPT family~\cite{DBLP:journals/corr/abs-2405-15189, shivashankar2024betterpythonprogrammingall, van2024llamea}, LLaMA family~\cite{DBLP:conf/euromlsys/GrubisicSSLMC24, DBLP:journals/corr/abs-2309-07062, li2024instructcoderinstructiontuninglarge}, Claude family~\cite{han2024generating, DBLP:journals/corr/abs-2405-15189}, and other open-source LMs~\cite{xu2024large, zhang2024codedpo}. For example,~\citet{han2024generating} leveraged Claude-3-haiku due to its ability to provide a robust semantic understanding and efficient processing of large-scale codebases, and~\citet{DBLP:journals/corr/abs-2404-18864} selected Gemini-Pro-1.0 to generate
synthetic code snippets based on its outstanding performance among other LMs in their empirical experiments. 

Despite not being explicitly trained for coding tasks, these general-purpose LMs were frequently chosen for code optimization due to several key factors: (1) They benefit from an extensive training on diverse datasets, leading to a more comprehensive understanding of language and improved contextual awareness~\cite{zhao2023survey}; (2) Their versatility allows them to handle a broader range of tasks beyond just generating optimized code, e.g., user query analysis~\cite{sun2024autosat}, and self-reflective evaluation~\cite{sun2024autosat, DBLP:conf/nips/MadaanTGHGW0DPY23}, making them suitable for more roles in the code optimization pipeline. 

\subsubsection{Code-specialized LMs}
A total of 43 code-specialized LMs were utilized in the reviewed studies, which are tailored specifically for code-related tasks, offering enhanced performance due to their targeted training on programming-specific datasets. The most widely used examples included Code LLaMA (11 times)~\cite{DBLP:journals/corr/abs-2407-02524, gao2024searchbasedllmscodeoptimization,   DBLP:journals/corr/abs-2405-15189}, DeepSeekCoder (seven times)~\cite{DBLP:journals/corr/abs-2406-11935, DBLP:journals/corr/abs-2407-14044, zhang2024codedpo}, StarCoder (four times)~\cite{DBLP:journals/corr/abs-2405-15189,   DBLP:journals/nature/RomeraParedesBNBKDREWFKF24,   DBLP:journals/corr/abs-2407-14044}, and CodeT5 (four times)~\cite{pan2024measuringcodeefficiencyoptimization,   DBLP:journals/corr/abs-2312-05657, DBLP:journals/pacmpl/YeZSS23}. For instance,~\citet{li2024instructcoderinstructiontuninglarge} opted for Code LLaMA as it combines the foundational strengths of LLaMA-2 with code-specific adjustments, allowing for better performance in coding tasks, and~\citet{huang2024effi} chose DeepSeek-Coder due to its strong performance on existing coding benchmarks and its accessibility for further fine-tuning. 

Compared to general-purpose LMs, these code-specialized models have several advantages: (1) They are designed to better capture code semantics like dependencies, function calls, and complex control flows, resulting in a better understanding of the code structures and subtle semantics~\cite{li2024instructcoderinstructiontuninglarge, pan2024measuringcodeefficiencyoptimization}; (2) As shown in Table~\ref{tb:base_lms}, they are typically smaller and open-source, enabling easier fine-tuning for specific tasks like code repair, completion, and translation~\cite{DBLP:journals/concurrency/HanLDLZ24}. Yet, they may lack the versatility of general-purpose LMs.

\subsubsection{Transformer LMs}
Finally, two representative foundational LMs, both built upon basic Transformer architectures, were utilized. Specifically,~\citet{pan2024ecodemasteringefficientcode} employed multiple instances of BERT-tiny---a small Transformer with bidirectional attention and 4.4 million parameters---to extract features tailored to different input types due to their computational efficiency, and~\citet{DBLP:journals/corr/abs-2109-13498} used a standard Transformer structure to reduce computational burden for instruction-level code optimization. Although smaller and less accurate than their successors, these foundational models are essential for applications requiring quick response times and lower computational costs. Because they have low computational requirements, they can run on developer PCs or a local cluster without sending the data to a remote cloud server. This makes them attractive to companies who do not want to send data and code to untrusted service providers. 

\keybox{
\faIcon{search} \textit{\textbf{Finding \thefindingcount:} General-purpose LMs were the most widely used for code optimization due to their broad understanding and reasoning capabilities. Code-specialized LMs excel in targeted optimization but may lack versatility. Meanwhile, foundational transformer-based LMs, though less accurate, remain crucial for resource-intensive applications.}
\addtocounter{findingcount}{1}
}

\suggestbox{
\faIcon{thumbs-up} \textit{\textbf{Recommendation \thesuggestioncount:} Given the summaries of characteristics of different LMs in Table~\ref{tb:base_lms}, future studies can select the most suitable LMs based on their needs, or explore integrated workflows where different model types are combined to maximize their complementary strengths.}
\addtocounter{suggestioncount}{1}
}

\subsection{RQ1.2: What Were Their Sizes?}
\label{subssec:LM_sizes}
In this subsection, we explore the parameter sizes of the foundation LMs used for code optimization, as depicted in Figure~\ref{fig:parameter_size}. Following the definitions of different scales of LMs introduced by~\citet{minaee2024large}, we categorized them into four distinct sizes based on their parameter counts: (1) \emph{Very large:} over 100 billion; (2) \emph{Large:} 10 to 100 billion; (3) \emph{Medium:} 1 to 10 billion; and (4) \emph{Small:} up to 1 billion. This categorization sheds light on the scale and capabilities of different LMs, ranging from lightweight, efficient ones to highly complex ones capable of advanced tasks.

\subsubsection{Very large models.} Notably, 49 very large models were utilized in the primary studies, ranging from 175 billion parameters (e.g., GPT-3.5) to 540 billion (e.g., Gemini-Pro) and even 1.8 trillion (e.g., GPT-4)\footnote{Given the significant gap between 540B and 1.8T parameters, we divide this category into two separate bars in Figure~\ref{fig:parameter_size}.}, making them capable of handling highly intricate code optimization tasks, e.g., code performance prediction~\cite{DBLP:journals/corr/abs-2406-04693, DBLP:conf/nips/MadaanTGHGW0DPY23, DBLP:conf/nips/ShinnCGNY23, peng2024large}, code translation~\cite{DBLP:journals/concurrency/HanLDLZ24, DBLP:journals/corr/abs-2408-03408}, and code vectorization~\cite{DBLP:journals/corr/abs-2406-04693}. An illustrative example was provided by~\citet{sun2024autosat}, where GPT-4 was combined with an agentic workflow, serving as a task advisor, code optimizer, and performance evaluator, delivering robust performance across multiple complex optimization roles.

\subsubsection{Large models.} Large models, encompassing 12 billion to 70 billion or more parameters, offered deeper contextual understanding and were employed in 39 instances for advanced scenarios, such as learning from fast-slow code pairs~\cite{DBLP:journals/corr/abs-2406-11935, DBLP:conf/iclr/ShypulaMZ0GYHNR24}, problem reasoning~\cite{DBLP:journals/corr/abs-2401-08500, li2024instructcoderinstructiontuninglarge}, and decoding code representations~\cite{pan2024ecodemasteringefficientcode}. For example,~\citet{DBLP:journals/corr/abs-2401-08500} leveraged DeepSeek-33B for reasoning about coding problems and generating edge test cases, as it offers a superior understanding of code contexts and effectively optimizes complex programs.

\subsubsection{Medium models.} On the other hand, medium-sized models, with 1 billion to 8 billion parameters, appeared 35 times and provided a balance of performance and resource use, making them suitable for moderately complex tasks like test-case generation~\cite{huang2024effi}, initial (slow) code generation~\cite{zhang2024codedpo, DBLP:journals/corr/abs-2406-12502}, and optimization pass sampling~\cite{DBLP:conf/euromlsys/GrubisicSSLMC24, DBLP:journals/corr/abs-2403-14714, DBLP:journals/corr/abs-2407-02524}. For instance,~\citet{DBLP:conf/euromlsys/GrubisicSSLMC24} utilized LLaMA-2-7B to generate optimization passes for code based on the input it received, leveraging its learned knowledge to produce effective sequences of transformations, which achieved a good balance between optimization ability and computational efficiency.

\begin{figure}[!t]
\centering
\begin{minipage}{0.39\textwidth}
\begin{tikzpicture}
\begin{axis}[
    xbar=0.05cm,
    width=10cm,
    height=6cm,
    xmax=48,
    bar width=0.4cm,
    bar shift=0cm, 
 enlarge y limits=0.2,
 enlarge x limits=0.02,
    legend style={at={(0.8,0.9)},anchor=north,legend columns=2,font=\small},
    legend cell align={left},
    legend image code/.code={
    \draw [#1] (0cm,-0.1cm) rectangle (0.15cm,0.15cm); },
   xlabel={\# studies},
     x label style={at={(0.5,0.02)},font=\Large},
     nodes near coords,
          point meta=explicit symbolic,
          every node near coord/.append style={anchor=west,font=\large},
        symbolic y coords={A,B,C,D,E,F},
         y tick label style={font=\large},
          x tick label style={font=\large},
          yticklabels={,,,,},
         ytick=data
    ]

\addplot[fill=frenchblue!50, draw=black] coordinates {(13,B)[13 (4.4M-770M)]};
\addlegendentry{Small} 
\addplot[fill=iceberg!50, draw=black] coordinates {(35,C)[35 (1B-8B)]};
\addlegendentry{Medium} 
\addplot[fill=cerublue!70, draw=black] coordinates {(39,D)[39 (12B-70B)]};
\addlegendentry{Large} 
\addplot[fill=red!60, draw=black, forget plot] coordinates {(23,E) [23 (175B-540B)]};
\addplot[fill=red!60, draw=black] coordinates {(26,F) [26 (1.8T)]};
\addlegendentry{Very large} 

\end{axis}
\end{tikzpicture}
\vspace{-0.7cm}
  \caption{Distribution of parameter sizes (one study might be in multiple categories).}
 \label{fig:parameter_size}
  \end{minipage}
~\hspace{0cm}
\begin{minipage}{0.57\textwidth}
\centering
\begin{tikzpicture}[scale=0.8]
\tikzstyle{every node}=[font=\large]
\definecolor{beaublue}{rgb}{0.74, 0.83, 0.9}
\definecolor{applegreen}{rgb}{0.55, 0.71, 0.0}
\definecolor{dollarbill}{rgb}{0.52, 0.73, 0.4}
\definecolor{ao(english)}{rgb}{0.0, 0.5, 0.0}
\definecolor{amaranth}{rgb}{0.9, 0.17, 0.31}
\definecolor{cerublue}{rgb}{0.16, 0.32, 0.75}
\definecolor{frenchblue}{rgb}{0.0, 0.45, 0.73}
\definecolor{iceberg}{rgb}{0.44, 0.65, 0.82}
  
  
    \pie[
        /tikz/every pin/.style={align=center},
        text=pin,
        pin distance = 1,
        rotate=0,
        explode=0, 
        color={cerublue!70, iceberg!50, frenchblue!50},
        ]
        {
        57/\texttt{Leveraging off-the-shelf LMs} (30),
        43/\texttt{Pre-training \& fine-tuning} (23)
        }

    \end{tikzpicture}
~\vspace{-0.13cm}
\caption{Distribution of training the LMs.}
\label{fig:pretrain-finetune}
  \end{minipage}
 \end{figure}

\subsubsection{Small models.} Finally, 13 small models were employed in the primary studies, ranging from 4.4 million to 770 million parameters, which makes them suitable for basic yet crucial tasks, such as input preprocessing, natural language analysis, and type inference~\cite{DBLP:journals/pacmpl/YeZSS23, DBLP:journals/corr/abs-2109-13498, DBLP:conf/hpec/GuoM22, DBLP:conf/sigsoft/GargMCSW22}. An example of this would be~\citet{pan2024ecodemasteringefficientcode}, which used Bert-tiny (4.4M) for encoding multiple types of code contexts, as it provides sufficient accuracy without requiring extensive computational resources.

This taxonomy gives an overview of the LMs used for code optimization and highlights the importance of researchers choosing the correct language models for optimizing code effectively and efficiently, as different tasks may require models of varying magnitudes to achieve desired outcomes.
Additionally, we must emphasize that this taxonomy is provisional, as the definition of "large" will likely evolve over time. In other words, with the development of more powerful and efficient hardware and training paradigms, models with even more parameters may become standard, pushing the boundaries of a large language model. This evolving landscape will require continuous re-evaluation of the criteria used to define large and small models, reflecting the dynamic nature of AI development.

\keybox{
\faIcon{search} \textit{\textbf{Finding \thefindingcount:} The size of language models used for code optimization varies significantly depending on the specific task, with larger models generally being more popular.}
\addtocounter{findingcount}{1}
}

\suggestbox{
\faIcon{thumbs-up} \textit{\textbf{Recommendation \thesuggestioncount:} (1) It is crucial for researchers to carefully select language models suited to their specific optimization needs. (2) With the fast development of LMs, the standard of model parameters could evolve; therefore, researchers should be careful when using the term ``large''.}
\addtocounter{suggestioncount}{1}
}

\subsection{RQ1.3: How Were They Trained?}
\label{subsec:pretrain_finetune}
This sub-question examines the pre-training and fine-tuning processes of the language models, as shown in Figure~\ref{fig:pretrain-finetune}. Understanding these processes is important as they shed light on the methodologies to enhance model performance and tailor them to specific requirements.

\subsubsection{Leveraging off-the-shelf LMs}
DNNs need to learn from data. Model training is key to exposing LMs to vast amounts of data, enabling them to capture intricate patterns and relationships within the code. Many LMs are trained on diverse datasets, often including code, enabling them to acquire capabilities in code reasoning. This makes it feasible to utilize these off-the-shelf pre-trained models through techniques like prompt engineering \cite{DBLP:journals/corr/abs-2310-02304, DBLP:journals/corr/abs-2406-07496, gao2024searchbasedllmscodeoptimization} without the need for fine-tuning. As can be seen in Figure~\ref{fig:pretrain-finetune}, 57\% of the primary studies directly leveraged off-the-shelf pre-trained LMs~\cite{han2024generating, van2024llamea, jain2022jigsaw, palkowski2024gpt, sun2024autosat}. By leveraging open, pre-trained LMs, researchers can access large models without paying the overhead of training these models, which 
is typically beyond the reach of most academic institutions and individual researchers. However, relying solely on off-the-shelf LMs may lead to challenges such as potential biases embedded within pre-training datasets, limited adaptability to highly domain-specific tasks, and a lack of transparency in model behavior.

\subsubsection{Pre-training and fine-tuning}
In contrast, 23 studies (43\%) fine-tuned the LMs on their own program datasets, which aims to adapt pre-trained LMs to smaller and more focused requirements, thereby enhancing their accuracy and effectiveness in optimizing specialized code~\cite{xu2024code, pan2024ecodemasteringefficientcode, li2024instructcoderinstructiontuninglarge, DBLP:conf/iclr/ShypulaMZ0GYHNR24, DBLP:journals/corr/abs-2406-11935}. This distribution stresses the importance of fine-tuning as a crucial step in the workflow of leveraging LMs for code optimization. For instance,~\citet{DBLP:conf/iclr/ShypulaMZ0GYHNR24} collected an original dataset with slow and efficient code pairs, then fine-tuned their models on this dataset to allow the LM to generate more contextually relevant and accurate code improvements. Similarly, ~\citet{DBLP:journals/corr/abs-2407-02524} fine-tuned a pre-trained Code LLaMA on a vast corpus of LLVM-IR and assembly code, enhancing its understanding of compiler semantics and optimization techniques, which was further refined by instruction-tuning the model for LM-emulated compiler optimization tasks, yielding significant improvements in performance. Nonetheless, fine-tuning can sometimes lead to overfitting, where the model becomes too specialized in the fine-tuning task and loses its generalization ability. 

On the other hand, only two of the 23 studies attempted to train their own LMs from scratch using relatively small LMs---\citet{DBLP:conf/sigsoft/GargMCSW22} pre-trained a PyMT5 LM (340M) from scratch on both English and source code from open-source repositories, and a standard Transformer LM of 80M parameters was pre-trained by \citet{DBLP:journals/corr/abs-2109-13498} on a large dataset of over 1.61M open-source programs, which helps the model learn general programming patterns. This distribution suggests that training LMs may demand extensive hardware and substantial energy consumption. As a result, studies should carefully consider using off-the-shelf LMs or training LMs on their own, balancing the trade-offs between performance and resource efficiency~\cite{jain2022jigsaw, qu2024dynamic, DBLP:journals/corr/abs-2408-03408, DBLP:conf/nips/ShinnCGNY23}.

\keybox{
\faIcon{search} \textit{\textbf{Finding \thefindingcount:} While a majority of studies (57\%) relied on off-the-shelf pre-trained models to save time and computational resources, fine-tuning remained a vital step for 43\% of studies to ensure the models were well-adapted to the specific tasks at hand.}
\addtocounter{findingcount}{1}
}

\suggestbox{
\faIcon{thumbs-up} \textit{\textbf{Recommendation \thesuggestioncount:} Future researchers should carefully consider whether to pre-train or fine-tune models based on their specific requirements to balance between model performance and computing resources.}
\addtocounter{suggestioncount}{1}
}


\section{RQ2: How Were LMs Applied to Code Optimization Tasks?}
\label{sec:RQ2_LM_application}
Understanding how LMs are applied to code optimization tasks helps identify the unique challenges that researchers and developers encounter and provides insights into how these advanced models can be leveraged to overcome those obstacles. Therefore, this section explores the various ways LMs are employed in optimizing code, highlighting the challenges faced by the community, their corresponding solutions, and the roles of LMs in these solutions.

\begin{table}[t!]
  \caption{Distribution of addressed challenges (one study might be in multiple categories).}
\centering
\footnotesize
\begin{adjustbox}{width=\linewidth,center}
\begin{tabular}{lllll}
\toprule
\textbf{Category} & \textbf{Total \#} & \textbf{Challenge} & \textbf{\# studies} & \textbf{Reference} \\ \hline
\multirow{5}{*}{Performance} & \multirow{5}{*}{49} & Limitation of one-step   generation & 18 & \cite{pan2024measuringcodeefficiencyoptimization,   DBLP:journals/corr/abs-2406-07496, DBLP:journals/corr/abs-2109-13498,   DBLP:journals/corr/abs-2405-15189, DBLP:conf/nips/MadaanTGHGW0DPY23,   DBLP:conf/apsec/ShirafujiOSMW23, DBLP:journals/corr/abs-2401-08500,   DBLP:journals/corr/abs-2406-04693, DBLP:journals/corr/abs-2312-05657,   ishida2024langpropcodeoptimizationframework, hemberg2024evolving,   han2024generating, xu2024code, qu2024dynamic, li2024falcon, xu2024optimizing,   zhang2024revolve, peng2024perfcodegen} \\ \cline{3-5} 
 &  & Balancing   correctness and performance & 15 & \cite{pan2024measuringcodeefficiencyoptimization,   pan2024ecodemasteringefficientcode, DBLP:journals/corr/abs-2405-15189,   DBLP:journals/corr/abs-2404-18864, DBLP:journals/corr/abs-2406-11935,   DBLP:journals/corr/abs-2407-14044, DBLP:journals/corr/abs-2406-12502,   DBLP:journals/corr/abs-2406-04693, van2024llamea, palkowski2024gpt,   qu2024dynamic, peng2024large, huang2024effi, zhang2024codedpo, van2024loop} \\ \cline{3-5} 
 &  & Reliance   on human experts & 13 & \cite{DBLP:journals/corr/abs-2310-02304,   yao2024rtlrewritermethodologieslargemodels,   DBLP:journals/corr/abs-2408-03408, DBLP:journals/corr/abs-2306-17077,   DBLP:conf/sigsoft/GargMCSW22, DBLP:journals/corr/abs-2309-14846,   choi2024iterative, han2024generating, van2024llamea, palkowski2024gpt,   sun2024autosat, xu2024code, xu2024optimizing} \\ \cline{3-5} 
 &  & Poor   code maintainability & 2 & \cite{shivashankar2024betterpythonprogrammingall,   DBLP:journals/concurrency/HanLDLZ24} \\ \cline{3-5} 
 &  & Hardware-dependent   performance variability & 1 & \cite{DBLP:conf/iclr/ShypulaMZ0GYHNR24} \\ \hline
\multirow{6}{*}{Code} & \multirow{6}{*}{24} & Complexity of code & 10 & \cite{yao2024rtlrewritermethodologieslargemodels,   DBLP:journals/corr/abs-2406-07496, DBLP:conf/nips/MadaanTGHGW0DPY23,   DBLP:conf/apsec/ShirafujiOSMW23,   DBLP:journals/nature/RomeraParedesBNBKDREWFKF24,   DBLP:journals/corr/abs-2406-04693, DBLP:journals/corr/abs-2407-02524,   DBLP:conf/hpec/GuoM22, xu2024large, wei2024improving} \\ \cline{3-5} 
 &  & Limitation   on localized code modifications & 4 & \cite{gao2024searchbasedllmscodeoptimization,   DBLP:conf/iclr/ShypulaMZ0GYHNR24, pan2024ecodemasteringefficientcode,   DBLP:journals/corr/abs-2309-07062} \\ \cline{3-5} 
 &  & Incomplete   code representation & 4 & \cite{DBLP:journals/corr/abs-2403-14714,   DBLP:journals/corr/abs-2407-02524, jain2022jigsaw, li2024falcon} \\ \cline{3-5} 
 &  & Limited   exploration of low-level languages & 3 & \cite{DBLP:conf/hpec/GuoM22,   DBLP:journals/corr/abs-2407-02524, wei2024improving} \\ \cline{3-5} 
 &  & Limited   applicability to real-world code & 2 & \cite{DBLP:journals/corr/abs-2109-13498,   choi2024iterative} \\ \cline{3-5} 
 &  & Limited   representation of problems & 1 & \cite{DBLP:journals/corr/abs-2401-08500} \\ \hline
\multirow{7}{*}{Dataset} & \multirow{7}{*}{18} & Limited efficiency-related   datasets & 9 & \cite{pan2024measuringcodeefficiencyoptimization,   DBLP:journals/corr/abs-2404-18864, DBLP:journals/corr/abs-2312-05657,   DBLP:conf/iclr/ShypulaMZ0GYHNR24, DBLP:journals/corr/abs-2406-11935,   DBLP:journals/corr/abs-2407-14044, DBLP:conf/sigsoft/GargMCSW22, DBLP:journals/corr/abs-2406-12502,   huang2024effi} \\ \cline{3-5} 
 &  & Reliance   on manually labeled data & 3 & \cite{DBLP:conf/nips/MadaanTGHGW0DPY23,   DBLP:conf/nips/ShinnCGNY23, zhang2024codedpo} \\ \cline{3-5} 
 &  & Limited   low-level language datasets & 2 & \cite{DBLP:journals/corr/abs-2408-03408,   DBLP:journals/corr/abs-2407-02524} \\ \cline{3-5} 
 &  & Limited   real-world datasets & 1 & \cite{DBLP:journals/corr/abs-2109-13498} \\ \cline{3-5} 
 &  & Limited   code maintainability datasets & 1 & \cite{shivashankar2024betterpythonprogrammingall} \\ \cline{3-5} 
 &  & Limited   code editing datasets & 1 & \cite{li2024instructcoderinstructiontuninglarge} \\ \cline{3-5} 
 &  & Limited   type inference datasets & 1 & \cite{DBLP:journals/pacmpl/YeZSS23} \\ \hline
\multirow{9}{*}{LM} & \multirow{9}{*}{15} & Limited generalizability across   domains & 3 & \cite{DBLP:journals/corr/abs-2408-03408,   DBLP:journals/corr/abs-2309-14846, DBLP:journals/corr/abs-2306-17077} \\ \cline{3-5} 
 &  & Inefficiency   of querying LMs & 2 & \cite{DBLP:journals/corr/abs-2310-02304,   xu2024code} \\ \cline{3-5} 
 &  & Limitation   of sampling methods & 2 & \cite{DBLP:conf/euromlsys/GrubisicSSLMC24,   DBLP:journals/corr/abs-2401-08500} \\ \cline{3-5} 
 &  & High   cost of fine-tuning & 2 & \cite{DBLP:journals/corr/abs-2306-17077,   DBLP:journals/corr/abs-2312-05657} \\ \cline{3-5} 
 &  & Hallucination   Issues of LMs & 2 & \cite{sun2024autosat,   peng2024perfcodegen} \\ \cline{3-5} 
 &  & Sycophancies   of LMs & 1 & \cite{peng2024perfcodegen} \\ \cline{3-5} 
 &  & Inherent   randomness of LMs & 1 & \cite{yao2024rtlrewritermethodologieslargemodels} \\ \cline{3-5} 
 &  & Handling   multiple types of inputs & 1 & \cite{pan2024ecodemasteringefficientcode} \\ \cline{3-5} 
 &  & Limited   exploration of solution space & 1 & \cite{DBLP:journals/nature/RomeraParedesBNBKDREWFKF24} \\ \hline
Compiler & 3 & Limited optimization ability of   compilers & 3 & \cite{yao2024rtlrewritermethodologieslargemodels,   DBLP:journals/corr/abs-2309-14846, DBLP:journals/corr/abs-2309-07062}
\\
\bottomrule
\end{tabular}
\end{adjustbox}
\label{tb:challenges}
\end{table}

\subsection{RQ2.1: What Common Challenges in Code Optimization Were Addressed?}
\label{subsec:common_challenges}
Table~\ref{tb:challenges} summarizes the common challenges addressed in the primary studies. In the table, we classify the challenges into five main groups, which are essential for identifying recurring issues and guiding the development of effective strategies to enhance code optimization.

\subsubsection{Performance-related challenges}
The most common challenges were performance-related, occurring 49 times in total. Among others, 18 studies highlighted the \textbf{lack of performance feedback during one-step LM inference}~\cite{pan2024measuringcodeefficiencyoptimization,   DBLP:journals/corr/abs-2406-07496, DBLP:journals/corr/abs-2109-13498,   DBLP:journals/corr/abs-2401-08500,   DBLP:journals/corr/abs-2405-15189}. For example,~\citet{DBLP:conf/nips/MadaanTGHGW0DPY23} pointed out that LMs do not always generate the best output on their first try and often require feedback mechanisms to learn from their mistakes and improve over time, so they proposed a SELF-REFINE framework, which allows LMs to generate feedback on their own outputs, significantly improving their code reasoning ability through iterative performance enhancement and self-generated feedback loops.

Additionally, 15 studies focused on \textbf{balancing correctness with performance}~\cite{DBLP:journals/corr/abs-2406-12502,   DBLP:journals/corr/abs-2406-04693, van2024llamea, palkowski2024gpt,   qu2024dynamic}. As~\citet{DBLP:journals/corr/abs-2405-15189} mentioned, while LMs have shown impressive capabilities in generating code, the efficiency of this code is often suboptimal, leading to slower execution times and higher resource consumption. Meanwhile, the challenge of \textbf{reliance on human expertise} to optimize code was underscored in 13 studies~\cite{DBLP:journals/corr/abs-2408-03408, DBLP:journals/corr/abs-2306-17077,   DBLP:conf/sigsoft/GargMCSW22, DBLP:journals/corr/abs-2309-14846}, highlighting the resource-intensive nature of manual interpretation and refinement. Other performance-related issues included \textbf{poor code maintainability}~\cite{shivashankar2024betterpythonprogrammingall,   DBLP:journals/concurrency/HanLDLZ24}, and challenges related to \textbf{hardware-dependent performance variability}~\cite{DBLP:conf/iclr/ShypulaMZ0GYHNR24}. Particularly,~\citet{DBLP:conf/iclr/ShypulaMZ0GYHNR24} disclosed that measuring performance on different hardware can lead to inconsistent results, making it difficult to reliably evaluate the effectiveness of optimizations. To address this, they used hardware simulators like gem5 \cite{binkert2011gem5} to obtain deterministic runtime measurements. Unfortunately, accurate hardware simulations are usually orders of magnitude slower than running programs on real hardware \cite{rachuj2020impact}; therefore, using simulators limits the size and the scale of the programs to be targeted. 

\subsubsection{Code-related challenges}
Among our primary studies, 24 challenges addressed were related to the code itself. In particular, the \textbf{complexity of code syntax} posed a significant challenge for LMs, as outlined in 10 studies~\cite{DBLP:conf/apsec/ShirafujiOSMW23,   DBLP:journals/nature/RomeraParedesBNBKDREWFKF24,   DBLP:journals/corr/abs-2406-04693}. For example,~\citet{DBLP:journals/corr/abs-2407-02524} mentioned that the vast number of possible optimizations, nested structures, and their intricate interactions highly complicated the optimization process, hindering the models' ability to generate effective solutions. Subsequently, four studies discussed \textbf{limitations on localized code modifications}, where existing methods often focus on localized code modifications that may not effectively address deeper performance issues, emphasizing the need for algorithm and data structure-level optimizations~\cite{gao2024searchbasedllmscodeoptimization,   DBLP:conf/iclr/ShypulaMZ0GYHNR24, pan2024ecodemasteringefficientcode,   DBLP:journals/corr/abs-2309-07062}. As supported by~\citet{gao2024searchbasedllmscodeoptimization}, there is a lack of combinatorial optimizations involving multiple code segments that require different optimization strategies. 

Further, \textbf{incomplete code representation} was a challenge covered in another four studies, which can lead to a loss of critical information necessary for effective optimization~\cite{DBLP:journals/corr/abs-2403-14714, DBLP:journals/corr/abs-2407-02524, jain2022jigsaw, li2024falcon}. Then, three studies addressed \textbf{limited exploration of low-level languages} like assembly code, which are more verbose and contains less structural semantics than high-level languages, making them even harder to interpret and optimize~\cite{DBLP:conf/hpec/GuoM22, DBLP:journals/corr/abs-2407-02524, wei2024improving}. Lastly, two studies specifically addressed the challenge of \textbf{generalizing the optimization to real-world code} by establishing datasets with real-world programs~\cite{DBLP:journals/corr/abs-2109-13498, choi2024iterative}, and~\citet{DBLP:journals/corr/abs-2401-08500} highlighted the difficulty of \textbf{representing complex problems}, since real-world code optimization problems are often nuanced and described in lengthy natural language specifications.

\subsubsection{Dataset-related challenges}
The limited availability of different datasets was the third major concern, being highlighted 18 times. Among these, the \textbf{lack of efficiency-related datasets} was the most common challenge, appearing in nine studies~\cite{DBLP:journals/corr/abs-2407-14044, DBLP:conf/sigsoft/GargMCSW22, DBLP:journals/corr/abs-2406-12502, huang2024effi}. For instance,~\citet{huang2024effi} mentioned that existing datasets often lack the necessary performance profiling data to evaluate and improve code efficiency, highlighting the need for high-quality datasets that focus on both correctness and efficiency metrics. Meanwhile, three studies noted the \textbf{dependence on manually labeled data}, which can be costly and time-consuming to obtain, limiting the scalability of code optimization methods~\cite{DBLP:conf/nips/MadaanTGHGW0DPY23, DBLP:conf/nips/ShinnCGNY23, zhang2024codedpo}. Moreover, the \textbf{lack of low-level language datasets} were each noted in two studies~\cite{DBLP:journals/corr/abs-2408-03408, DBLP:journals/corr/abs-2407-02524}, and one study each noted the need for specific datasets in four different areas, including code editing and type inference et al.~\cite{DBLP:journals/corr/abs-2109-13498, shivashankar2024betterpythonprogrammingall, li2024instructcoderinstructiontuninglarge, DBLP:journals/pacmpl/YeZSS23}.

\subsubsection{LM-related challenges}
The challenges related to LMs were mentioned 15 times, with \textbf{generalizability across domains} being the most frequently cited one~\cite{DBLP:journals/corr/abs-2408-03408, DBLP:journals/corr/abs-2309-14846, DBLP:journals/corr/abs-2306-17077}. As~\citet{DBLP:journals/corr/abs-2309-14846} reported, existing optimization techniques may not generalize well across different programming styles and patterns, creating a gap in their applicability. It was also found challenging to \textbf{handle the hallucination issues of LMs} (where LMs can generate outputs that are factually incorrect or nonsensical, potentially misleading users and reducing the reliability of the models~\cite{sun2024autosat, peng2024perfcodegen}); to \textbf{efficiently query LMs}~\cite{DBLP:journals/corr/abs-2310-02304, xu2024code}; to \textbf{sample candidates generated by the LMs}~\cite{DBLP:conf/euromlsys/GrubisicSSLMC24, DBLP:journals/corr/abs-2401-08500}; and to \textbf{fine-tune models with limited cost}~\cite{DBLP:journals/corr/abs-2306-17077, DBLP:journals/corr/abs-2312-05657}, as observed in two studies each. For instance,~\citet{DBLP:journals/corr/abs-2310-02304} identified that most existing methods focus on manually optimizing prompts or outputs, which can be inefficient and unscalable, so they designed a self-improving scaffolding program to structure interactions with LMs, strengthening the querying efficiency.

Moreover, one study each highlighted the following challenges: the \textbf{sycophancies of LMs}, where LMs may overly conform to user prompts without critical evaluations \cite{peng2024perfcodegen}; the \textbf{inherent randomness of LMs}, where the stochastic nature of model responses can lead to inconsistent outputs, posing a challenge in obtaining stable and predictable results for code optimization tasks~\cite{yao2024rtlrewritermethodologieslargemodels}; the difficulty of \textbf{handling multiple types of inputs}, where LMs struggle to process and integrate various data formats and input types simultaneously, limiting the code contexts ~\cite{pan2024ecodemasteringefficientcode}; and the \textbf{limited exploration of the solution space}, where LMs may not thoroughly investigate all possible optimization strategies, leading to suboptimal performance improvements~\cite{DBLP:journals/nature/RomeraParedesBNBKDREWFKF24}.

\subsubsection{Code-analysis-related challenges}
Finally, the \textbf{restricted abilities of code analysis tools} like compilers were covered in three studies, suggesting potential areas where LMs could enhance code analysis capabilities. For example,~\citet{yao2024rtlrewritermethodologieslargemodels} found that existing compiler-based approaches are often inadequate for handling complex Register Transfer Level (RTL) patterns. Similarly,~\citet{DBLP:journals/corr/abs-2309-14846} and~\citet{DBLP:journals/corr/abs-2309-07062} argued that while compilers can perform a range of automated optimizations, they may not capture high-level optimizations that require a deeper understanding of the code's logic and context, such as refactoring inefficient algorithms or modifying data structures. Yet, with LMs, it becomes possible to bridge this gap by leveraging their advanced semantic understanding and contextual analysis capabilities.

\keybox{
\faIcon{search} \textit{\textbf{Finding \thefindingcount:} The most commonly highlighted challenges were related to performance and code, including limitation of one-step optimization (18 studies), balancing correctness and efficiency (15 studies), and complexity of code syntax (10 studies).}
\addtocounter{findingcount}{1}
}

\suggestbox{
\faIcon{thumbs-up} \textit{\textbf{Recommendation \thesuggestioncount:} Attention should be given not only to common challenges but also to less frequently addressed issues that could be equally critical, such as the hallucination of LMs.}
\addtocounter{suggestioncount}{1}
}

\subsection{RQ2.2: How Were the Challenges Addressed Using LMs?}
\label{subsec:code_optimization_methods}
This subsection examines existing applications of LMs to address code optimization challenges, which can be broadly grouped into three categories: building specific models, leveraging prompt engineering, and formulating new code optimization problems, as illustrated in Table~\ref{tb:code_optimization}. By analyzing these techniques and their targeted challenges, researchers and practitioners can gain valuable insights into different methods, thereby making informed decisions when selecting approaches that align with their specific constraints and requirements, such as computational resources, task complexity, or data availability.

\begin{table}[t!]
  \caption{Distribution of code optimization techniques (one study might be in multiple categories).}
\centering
\footnotesize
\begin{adjustbox}{width=\linewidth,center}
\begin{tabular}{llp{3cm}lp{2.8cm}p{10cm}}
\toprule
\textbf{Category} & \textbf{Total \#} & \textbf{Technique} & \textbf{\# studies} & \textbf{Reference} & \textbf{Addressed challenge (\# studies)} \\ \hline
\multirow{7}{*}{Model-based} & \multirow{7}{*}{51} & Feedback-based iterative   optimization & 35 & \cite{yao2024rtlrewritermethodologieslargemodels,   DBLP:journals/corr/abs-2310-02304, DBLP:journals/corr/abs-2408-03408,   DBLP:journals/corr/abs-2406-07496, gao2024searchbasedllmscodeoptimization,   DBLP:journals/corr/abs-2405-15189, DBLP:conf/nips/MadaanTGHGW0DPY23,   DBLP:conf/nips/ShinnCGNY23, DBLP:conf/apsec/ShirafujiOSMW23,   DBLP:journals/nature/RomeraParedesBNBKDREWFKF24,   DBLP:journals/corr/abs-2401-08500, DBLP:journals/corr/abs-2404-18864,   DBLP:journals/corr/abs-2406-04693, DBLP:journals/corr/abs-2312-05657,   ishida2024langpropcodeoptimizationframework,   DBLP:journals/corr/abs-2407-14044, DBLP:journals/concurrency/HanLDLZ24,   DBLP:journals/corr/abs-2403-14714, DBLP:journals/corr/abs-2406-12502,   choi2024iterative, hemberg2024evolving, han2024generating, van2024llamea,   jain2022jigsaw, palkowski2024gpt, xu2024large, wei2024improving, xu2024code,   qu2024dynamic, li2024falcon, peng2024large, huang2024effi, zhang2024codedpo,   zhang2024revolve, van2024loop} & Limitation of one-step   optimization (14), Balancing correctness and performance (12), Complexity of   code (8), Reliance on human experts (8), Limited efficiency-related datasets   (5), Reliance on manually labeled data (4), Inefficiency of querying LMs (3),   Incomplete code representation (3), Hallucination Issues of LMs (2), High   cost of fine-tuning (1), Inherent randomness of LMs (1), Limited   generalizability across domains (1), Limited exploration of solution space   (1) \\ \cline{3-6} 
 &  & Agentic   workflow & 6 & \cite{DBLP:conf/nips/ShinnCGNY23,   wei2024improving, peng2024large, DBLP:journals/corr/abs-2406-04693,   sun2024autosat, zhang2024revolve} & Balancing correctness and   performance (2), Complexity of code (2), Reliance on human experts (1),   Reliance on manually labeled data (1), Hallucination Issues of LMs (1) \\ \cline{3-6} 
 &  & Compiler   emulation & 4 & \cite{DBLP:conf/hpec/GuoM22,   DBLP:journals/corr/abs-2407-02524, DBLP:journals/corr/abs-2309-07062,   DBLP:journals/corr/abs-2403-14714} & Complexity of code (2),   Incomplete code representation (2), Limited exploration of low-level   languages (2), Limited low-level language datasets (1), Limitation on   localized code modifications (1), Limited optimization ability of compilers   (1) \\ \cline{3-6} 
 &  & Direct   preference optimization & 3 & \cite{zhang2024codedpo,   DBLP:journals/corr/abs-2404-18864, DBLP:journals/corr/abs-2406-12502} & Balancing correctness and   performance (3), Limited efficiency-related datasets (2), Reliance on   manually labeled data (1) \\ \cline{3-6} 
 &  & Compiler   passes sampling & 1 & \cite{DBLP:conf/euromlsys/GrubisicSSLMC24} & Limitation of sampling methods   (1) \\ \cline{3-6} 
 &  & Ensemble   learning & 1 & \cite{DBLP:journals/corr/abs-2406-11935} & Balancing correctness and   performance (1), Limited efficiency-related datasets (1) \\ \cline{3-6} 
 &  & Encoder-decoder & 1 & \cite{pan2024ecodemasteringefficientcode} & Limitation on localized code   modifications (1), Handling multiple types of inputs (1) \\ \hline
\multirow{5}{*}{Prompt engineering} & \multirow{5}{*}{34} & Few-shot prompting & 11 & \cite{DBLP:conf/nips/MadaanTGHGW0DPY23,   DBLP:conf/apsec/ShirafujiOSMW23, DBLP:conf/iclr/ShypulaMZ0GYHNR24,   DBLP:journals/corr/abs-2407-14044, hemberg2024evolving, xu2024optimizing,   zhang2024codedpo, li2024instructcoderinstructiontuninglarge, DBLP:conf/nips/ShinnCGNY23,   DBLP:journals/nature/RomeraParedesBNBKDREWFKF24, van2024loop} & Limitation of one-step   optimization (3), Complexity of code (3), Reliance on manually labeled data   (3), Balancing correctness and performance (3), Limited efficiency-related   datasets (2), Reliance on human experts (1), Inefficiency of querying LMs (1) \\ \cline{3-6} 
 &  & Contextual   prompting & 9 & \cite{DBLP:journals/corr/abs-2408-03408,   jain2022jigsaw, xu2024large, wei2024improving, li2024falcon,   shivashankar2024betterpythonprogrammingall,   DBLP:journals/corr/abs-2405-15189, DBLP:journals/pacmpl/YeZSS23,   peng2024perfcodegen} & Limitation of one-step   optimization (2), Complexity of code (2), Incomplete code representation (2),   Poor code maintainability (1), Limited generalizability across domains (1) \\ \cline{3-6}
 &  & Chain-of-thought & 8 & \cite{gao2024searchbasedllmscodeoptimization,   DBLP:conf/nips/ShinnCGNY23, DBLP:conf/iclr/ShypulaMZ0GYHNR24,   DBLP:journals/corr/abs-2406-11935, sun2024autosat, xu2024code,   DBLP:journals/corr/abs-2406-07496,   ishida2024langpropcodeoptimizationframework} & Limitation of one-step   optimization (3), Limitation on localized code modifications (2), Reliance on   human experts (2), Balancing correctness and performance (1), Complexity of   code (1), Reliance on manually labeled data (1), Inefficiency of querying LMs   (1), Hallucination Issues of LMs (1) \\ \cline{3-6} 
 &  & Retrieval-augmented generation & 5 & \cite{gao2024searchbasedllmscodeoptimization,   DBLP:journals/corr/abs-2306-17077, DBLP:conf/iclr/ShypulaMZ0GYHNR24,   yao2024rtlrewritermethodologieslargemodels, xu2024optimizing} & Limitation on localized code   modifications (2), Reliance on human experts (2), Limitation of one-step   optimization (1), Hardware-dependent performance variability (1), High cost   of fine-tuning (1), Limited generalizability across domains (1) \\ \cline{3-6} 
 &  & Scaffolding   optimization & 1 & \cite{DBLP:journals/corr/abs-2310-02304} & Inefficiency of querying LMs   (1), Reliance on human experts (1) \\ \hline
\multirow{7}{*}{Problem formulation} & \multirow{7}{*}{33} & Dataset & 19 & \cite{pan2024measuringcodeefficiencyoptimization,   shivashankar2024betterpythonprogrammingall,   DBLP:journals/corr/abs-2309-14846, DBLP:journals/corr/abs-2109-13498,   li2024instructcoderinstructiontuninglarge,   DBLP:conf/euromlsys/GrubisicSSLMC24, DBLP:journals/corr/abs-2404-18864,   DBLP:conf/iclr/ShypulaMZ0GYHNR24, DBLP:journals/corr/abs-2309-07062,   DBLP:journals/corr/abs-2406-11935, DBLP:journals/corr/abs-2407-14044,   DBLP:conf/sigsoft/GargMCSW22, DBLP:journals/pacmpl/YeZSS23,   DBLP:journals/corr/abs-2403-14714, DBLP:journals/corr/abs-2406-12502,   DBLP:journals/corr/abs-2407-02524, xu2024code, huang2024effi,   zhang2024codedpo} & Limited efficiency-related   datasets (8), Balancing correctness and performance (7), Limitation of   one-step optimization (2), Limitation on localized code modifications (2),   Reliance on human experts (2), Incomplete code representation (2), Limited   low-level language datasets (1), Limited real-world datasets (1), Limited   code maintainability datasets (1), Limited code editing datasets (1), Limited   type inference datasets (1), Complexity of code (1), Reliance on manually   labeled data (1) \\ \cline{3-6} 
 &  & Reinforcement   learning & 6 & \cite{DBLP:journals/corr/abs-2404-18864,   DBLP:journals/corr/abs-2312-05657, han2024generating, li2024falcon,   DBLP:conf/nips/ShinnCGNY23, ishida2024langpropcodeoptimizationframework} & Limitation of one-step   optimization (3), Limited efficiency-related datasets (2), Balancing   correctness and performance (1), Reliance on manually labeled data (1),   Incomplete code representation (1), High cost of fine-tuning (1) \\ \cline{3-6} 
 &  & Search-based & 4 & \cite{gao2024searchbasedllmscodeoptimization,   DBLP:journals/corr/abs-2109-13498,   DBLP:journals/nature/RomeraParedesBNBKDREWFKF24, hemberg2024evolving} & Limitation of one-step   optimization (1), Complexity of code (1), Limitation on localized code   modifications (1), Limited exploration of solution space (1) \\ \cline{3-6} 
 &  & Code   token tree & 1 & \cite{qu2024dynamic} & Limitation of one-step   optimization (1), Balancing correctness and performance (1) \\ \cline{3-6} 
 &  & Modular   generation & 1 & \cite{xu2024large} & Complexity of code (1) \\ \cline{3-6} 
 &  & Metric   design & 1 & \cite{pan2024measuringcodeefficiencyoptimization} & Limitation of one-step   optimization (1), Balancing correctness and performance (1) \\ \cline{3-6} 
 &  & Diff   synthesis & 1 & \cite{DBLP:journals/corr/abs-2309-14846} & Reliance on human experts (1),   Limited generalizability across domains (1)
\\
\bottomrule
\end{tabular}
\end{adjustbox}
\label{tb:code_optimization}
\end{table}

\subsubsection{Model-based techniques}
The most common strategy focused on designing specialized models to directly tackle code optimization challenges, with 51 instances in total. 
In particular, \textbf{feedback-based iterative techniques} were the most prominent in this category, utilized in 35 studies~\cite{DBLP:journals/concurrency/HanLDLZ24,   DBLP:journals/corr/abs-2403-14714, DBLP:journals/corr/abs-2406-12502,   choi2024iterative, hemberg2024evolving}. These methods employ various kinds of feedback information from evaluations to effectively address a range of challenges, where the most frequent ones included the limitation of one-step optimization (14 studies), balancing correctness and performance (12 studies), the complexity of code (eight studies), and reliance on human experts (eight studies). For example,~\citet{DBLP:journals/corr/abs-2312-05657} proposed PerfRL, which integrates LMs with a reinforcement learning framework that utilizes compilation and runtime feedback derived from unit testing to iterative optimize code runtime efficiency.~\citet{peng2024large} utilized an agentic workflow to optimize the energy consumption of code, including a generator LM agent to generate and iteratively optimize code, and an evaluator LM agent to provide feedback on correctness and energy consumption. Moreover, it is also feasible to use the same LM for both code optimization and performance evaluation, as demonstrated by the SELF-REFINE model~\cite{DBLP:conf/nips/MadaanTGHGW0DPY23}. However, these feedback-based techniques can be computationally intensive and may require significant resources to generate and process feedback.

\textbf{Agentic approaches} were explored in six studies, primarily targeting challenges like balancing correctness and performance~\cite{peng2024large, DBLP:journals/corr/abs-2406-04693}, the complexity of code~\cite{wei2024improving, DBLP:journals/corr/abs-2406-04693}, and the hallucination issues of LMs~\cite{sun2024autosat}. For example,~\citet{sun2024autosat} presented AutoSAT, a framework that leverages LM agents as (1) an actor that makes decisions and actions, (2) a code optimizer that generates optimized codes based on feedback, and (3) an evaluator that provides feedback based on existing code, thereby, this approach helps validate and refine the heuristics generated by LMs, reducing the impact of hallucinations and ensuring more reliable outputs. Noteworthy, these approaches may struggle with scalability and the complexity of coordinating multiple agents.

Another four studies designed \textbf{compiler emulation} models to address challenges such as the complexity and limited exploration of code~\cite{DBLP:conf/hpec/GuoM22, DBLP:journals/corr/abs-2407-02524}, incomplete code representation~\cite{DBLP:journals/corr/abs-2407-02524, DBLP:journals/corr/abs-2403-14714}, and the restricted optimization capabilities of traditional compilers~\cite{DBLP:journals/corr/abs-2309-07062}. Specifically, in~\citet{DBLP:journals/corr/abs-2309-07062}, the LM was implemented to not only suggest optimization pass lists but also to generate optimized code directly as a compiler, allowing the model to bypass traditional compilation processes and mitigate the challenge of needing to compile multiple times to evaluate different optimization strategies. These models, however, may require domain-specific knowledge and datasets.

Other techniques included \textbf{direct preference optimization (DPO)}, which involves training the LMs to rank and select optimal outputs based on pre-defined criteria, thereby enhancing the overall optimization quality~\cite{zhang2024codedpo, DBLP:journals/corr/abs-2404-18864, DBLP:journals/corr/abs-2406-12502}; \textbf{compiler passes sampling}, where a deterministic sampling technique is implemented to utilize LMs for structured explorations of optimization passes~\cite{DBLP:conf/euromlsys/GrubisicSSLMC24}; \textbf{ensemble learning}, which merges the parameters or outputs of multiple LMs to create a single unified model~\cite{DBLP:journals/corr/abs-2406-11935}; and \textbf{encoder-decoder models}, which consists of multiple BERT-tiny encoders and a GPT-NEO decoder to handle distinct code contexts effectively~\cite{pan2024ecodemasteringefficientcode}.

\subsubsection{Prompt engineering techniques}
By carefully designing and structuring input prompts to achieve desired outputs, prompt engineering formed the second major category. Various prompting techniques are used to guide LMs for code optimization, explored in 34 instances. Firstly, \textbf{few-shot prompting}, used in 11 studies, involves prompting an LM with a few example inputs and outputs to enable it to generalize to new optimization tasks, addressing challenges like the limitation of one-step optimization~\cite{DBLP:conf/nips/MadaanTGHGW0DPY23, DBLP:conf/apsec/ShirafujiOSMW23, xu2024optimizing} and reliance on manually labeled data~\cite{DBLP:conf/nips/MadaanTGHGW0DPY23, zhang2024codedpo, DBLP:conf/nips/ShinnCGNY23}. For example, when prompting the LM,~\citet{DBLP:journals/nature/RomeraParedesBNBKDREWFKF24} constructed prompts by combining several best-performed programs from the program database to enable the LM to learn efficient patterns and generalize them. One potential challenge is these techniques may be limited by the quality and representativeness of the few examples provided.

\textbf{Contextual prompting}, used in nine studies, targets challenges such as the complexity of code~\cite{xu2024large, wei2024improving} and incomplete code representation~\cite{jain2022jigsaw, li2024falcon}. It involves providing models with comprehensive and relevant contextual information to improve their understanding of the code to optimize, often in the format of a prompt template. For example,~\citet{DBLP:journals/corr/abs-2405-15189} prompted the LM with multiple types of contexts, including task description, test cases, initial code, overhead analysis, and optimization rules, thus improving its performance and efficiency in generating high-quality outputs. However, there is a risk of overwhelming the model with too much information, which can lead to a decrease in accuracy.

\textbf{Chain-of-thought (CoT) prompting}, adopted in eight studies, improves the reasoning capabilities of LMs in complex code optimization tasks by guiding the model to generate intermediate reasoning steps before arriving at the final answer. Challenges addressed included the limitation of one-step optimization~\cite{xu2024code, DBLP:journals/corr/abs-2406-07496, ishida2024langpropcodeoptimizationframework} and limitations on localized code modifications~\cite{gao2024searchbasedllmscodeoptimization, DBLP:conf/iclr/ShypulaMZ0GYHNR24}. In particular,~\citet{xu2024code} proposed a novel self-checking code-CoT approach, where the LM is prompted to decompose the code optimization task into logical steps, generate test cases, self-check, and iteratively refine the code to ensure the performance and correctness of the final code. Notably, these methods may require more computational resources and careful design of intermediate steps.

\textbf{Retrieval-augmented generation (RAG)}, explored in five studies, leverages external knowledge to address challenges like the limitation of localized code modifications~\cite{gao2024searchbasedllmscodeoptimization, DBLP:conf/iclr/ShypulaMZ0GYHNR24}, reliance on human experts~\cite{DBLP:journals/corr/abs-2306-17077, xu2024optimizing}, and high cost of fine-tuning~\cite{DBLP:journals/corr/abs-2306-17077}. As the example in~\citet{xu2024optimizing}, when a user inputs a piece of code along with optimization instructions, the system performs a query index search to match the input with the most relevant code embeddings from a customized codebase, which are then retrieved and integrated with the original prompts to guide the LM in generating better-optimized code. While contextual prompting also relies on pre-provided context to guide the LM, RAG can actively retrieve relevant data from external databases in real-time, providing more dynamic and up-to-date information. Yet, it can be challenging to select the most appropriate retrieval metrics and ensure the accuracy and relevance of the retrieved materials.

Finally,~\citet{DBLP:journals/corr/abs-2310-02304} opted for \textbf{scaffolding optimization}, where a scaffolding program that can prompt an LM in a structured way is utilized to refine itself through multiple iterations.

\subsubsection{Problem formulation-based techniques}
The last category involved formulating novel problems with new objectives to tackle foundational challenges in code optimization, as demonstrated 33 times. Among others, 19 studies opted for \textbf{dataset formulation}, where new types of code optimization datasets are collected, addressing issues related to incomplete datasets of different purposes, e.g., those for balancing correctness and efficiency~\cite{pan2024measuringcodeefficiencyoptimization, DBLP:journals/corr/abs-2404-18864, DBLP:conf/iclr/ShypulaMZ0GYHNR24}, low-level languages~\cite{DBLP:journals/corr/abs-2407-02524}, real-world scenarios~\cite{DBLP:journals/corr/abs-2109-13498}, type inference~\cite{DBLP:journals/pacmpl/YeZSS23}, and code editing~\cite{li2024instructcoderinstructiontuninglarge}. For example,~\citet{DBLP:conf/iclr/ShypulaMZ0GYHNR24} proposed PIE, a C++ dataset with over 77k pairs of competitive programming submissions, where each pair consists of a slower and a corresponding faster version of code from the same user, serving as a fundamental dataset for code optimization. Based on PIE,~\citet{DBLP:journals/corr/abs-2406-11935} introduced PIE-problem, which is supplemented with code pairs not only from the same user but also from the same coding problem, and it retains only pairs that have greater than 90\% relative runtime improvement.
For these techniques, ensuring the established datasets are high-quality and capture diverse code optimization scenarios can be a major concern.

Six studies transformed code optimization to a \textbf{reinforcement learning (RL) problem}, treating the task as a sequential decision-making process where the LM makes a series of optimization actions based on feedback from the environment, e.g., code execution~\cite{han2024generating, DBLP:journals/corr/abs-2312-05657, DBLP:journals/corr/abs-2404-18864, ishida2024langpropcodeoptimizationframework}, compiler analysis~\cite{li2024falcon, DBLP:journals/corr/abs-2312-05657}, and LM evaluator agent~\cite{DBLP:conf/nips/ShinnCGNY23}. Unlike search-based approaches, which explore a predefined search space for optimal solutions, RL continuously adapts its strategy through interactions with the code and execution environment. Yet, they face challenges including designing appropriate reward/feedback mechanisms and balancing between exploration and exploitation.

Likewise, four studies formulated code optimization as a \textbf{search-based problem}~\cite{gao2024searchbasedllmscodeoptimization, DBLP:journals/corr/abs-2109-13498, DBLP:journals/nature/RomeraParedesBNBKDREWFKF24, hemberg2024evolving}. These methods aim to conceptualize the task of code optimization to search for the most effective modifications within a vast space of potential solutions, allowing for a systematic exploration of the optimization space, and enabling the identification of complex optimization patterns that may not be captured through one-step optimization methods~\cite{DBLP:journals/corr/abs-2109-13498}. For instance,~\citet{gao2024searchbasedllmscodeoptimization} employed evolutionary search algorithms and execution feedback to guide the LMs in refining the optimization in a framework called Search-Based LLM (SBLLM). However, the complexity of the search space may make the problem computationally intensive, and the search process may be trapped in local optima, limiting the effectiveness of optimization.

Less frequent methods included \textbf{code token tree (CTT)}, a dynamic updating mechanism that guides the code generation process toward more optimized solutions by leveraging historical performance data to improve future code performance~\cite{qu2024dynamic}; \textbf{modular generation}, which generates modules on-the-fly to efficiently address code structure complexity~\cite{xu2024large}; \textbf{metric design}, where new evaluation metrics such as Normalized Performance Index (NPI) are designed to encourage LMs to prioritize efficiency in their outputs~\cite{pan2024measuringcodeefficiencyoptimization}; and \textbf{diff synthesis}, which regards optimized code as diff files with only minor code modifications, minimizing the risk of introducing bugs~\cite{DBLP:journals/corr/abs-2309-14846}.

\keybox{
\faIcon{search} \textit{\textbf{Finding \thefindingcount:} Model-based techniques were highly popular and effective (51 instances), but may face scalability issues. Prompt engineering methods are data-efficient, though their design depends on expert knowledge (34 instances). Meanwhile, problem formulation-based solutions offer flexibility in optimization but demand significant effort in problem and dataset preparation (33 instances).} 
\addtocounter{findingcount}{1}
}

\suggestbox{
\faIcon{thumbs-up} \textit{\textbf{Recommendation \thesuggestioncount:} (1) The results in Table~\ref{tb:code_optimization} can help identify the most suitable solutions for specific challenges at hand. (2) Researchers may design their own techniques by integrating insights from existing ones to address specific code optimization challenges effectively.}
\addtocounter{suggestioncount}{1}
}

\subsection{RQ2.3: What Were the Roles of LMs?}
\label{subsec:roles_LMs}
In this section, we review and summarize the key roles of LMs in the code optimization pipeline, as listed in Table~\ref{tb:role_LMs}. This question is essential for researchers and practitioners to understand how these LMs contribute to various stages of the optimization process and determine which other techniques to use alongside LMs to boost their performance. 

\subsubsection{Generation}
The generation category was the most widely studied, appearing in 73 instances across all primary studies. Notably, the most fundamental role in this category was \textbf{optimizer}, seen in 46 studies~\cite{choi2024iterative, hemberg2024evolving, han2024generating, van2024llamea, jain2022jigsaw}, where LMs are used directly to optimize existing poor-performance code, while the \textbf{generator} role focuses on generating initial code seeds based on natural language specifications, as used in 21 studies~\cite{DBLP:journals/corr/abs-2406-07496, DBLP:journals/corr/abs-2405-15189, DBLP:conf/nips/MadaanTGHGW0DPY23, DBLP:conf/nips/ShinnCGNY23}. Besides, four studies employed LMs to emulate \textbf{compilers} by learning compiler transformations and generating low-level code~\cite{DBLP:conf/hpec/GuoM22, DBLP:journals/corr/abs-2309-07062, DBLP:journals/corr/abs-2403-14714, DBLP:journals/corr/abs-2407-02524}, and one study each used LMs to \textbf{generate diff} files that representing code improvements~\cite{DBLP:journals/corr/abs-2309-14846} and \textbf{decode embeddings} from the encoder LMs and generate optimized code~\cite{pan2024ecodemasteringefficientcode}. Furthermore, these roles are often combined with various techniques to enhance their capabilities, including correctness analysis tools like compilers~\cite{choi2024iterative, palkowski2024gpt}, unit tests~\cite{xu2024large, xu2024code}, static analysis~\cite{peng2024large}, code profiling tools~\cite{DBLP:journals/corr/abs-2407-14044, DBLP:conf/iclr/ShypulaMZ0GYHNR24}, long/short term memory~\cite{li2024falcon, DBLP:conf/nips/ShinnCGNY23}, and retrieval knowledge bases~\cite{DBLP:journals/corr/abs-2306-17077}.

\subsubsection{Evaluation}
The evaluator role was explored in 10 studies, where LMs are prompted to assess the correctness, performance or quality of code, identify bugs, validate outputs, and ensure compliance with specifications~\cite{xu2024code, peng2024large,   pan2024ecodemasteringefficientcode, DBLP:journals/corr/abs-2406-07496}, providing insightful and explainable feedback. Moreover, they are often used together with code contexts like unit tests and problem descriptions to make accurate assessments~\cite{sun2024autosat}, and with code optimizers to reflect the feedback and iteratively refine code~\cite{DBLP:conf/nips/ShinnCGNY23}. While LM-based evaluators are often faster than traditional compilers, static analysis, and manual reviews, there could be potential accuracy and consistency issues due to the inherent hallucination nature of LMs~\cite{sun2024autosat}.

\begin{table}[t!]
  \caption{Distribution of roles of LMs (one study might be in multiple categories).}
\centering
\footnotesize
\begin{adjustbox}{width=\linewidth,center}
\begin{tabular}{llllp{12cm}}
\toprule
\textbf{Category} & \textbf{Total \#} & \textbf{Role} & \textbf{\# studies} & \textbf{Reference} \\ \hline
\multirow{5}{*}{Generation} & \multirow{5}{*}{73} & Optimizer & 46 & \cite{choi2024iterative,   hemberg2024evolving, han2024generating, van2024llamea, jain2022jigsaw,   palkowski2024gpt, sun2024autosat, xu2024large, wei2024improving, xu2024code,   qu2024dynamic, li2024falcon, peng2024large, huang2024effi, xu2024optimizing,   zhang2024codedpo, yao2024rtlrewritermethodologieslargemodels,   DBLP:journals/corr/abs-2310-02304, DBLP:journals/corr/abs-2408-03408,   pan2024measuringcodeefficiencyoptimization,   shivashankar2024betterpythonprogrammingall,   DBLP:journals/corr/abs-2406-07496, gao2024searchbasedllmscodeoptimization,   DBLP:journals/corr/abs-2109-13498, DBLP:journals/corr/abs-2405-15189,   DBLP:conf/nips/MadaanTGHGW0DPY23, li2024instructcoderinstructiontuninglarge,   DBLP:conf/nips/ShinnCGNY23, DBLP:conf/apsec/ShirafujiOSMW23, DBLP:conf/euromlsys/GrubisicSSLMC24,   DBLP:journals/nature/RomeraParedesBNBKDREWFKF24,   DBLP:journals/corr/abs-2401-08500, DBLP:journals/corr/abs-2306-17077,   DBLP:journals/corr/abs-2404-18864, DBLP:journals/corr/abs-2406-04693,   DBLP:journals/corr/abs-2312-05657, DBLP:conf/iclr/ShypulaMZ0GYHNR24,   ishida2024langpropcodeoptimizationframework,   DBLP:journals/corr/abs-2406-11935, DBLP:journals/corr/abs-2407-14044,   DBLP:journals/concurrency/HanLDLZ24, DBLP:conf/sigsoft/GargMCSW22,   DBLP:journals/corr/abs-2406-12502, zhang2024revolve, van2024loop,   peng2024perfcodegen} \\ \cline{3-5} 
 &  & Generator & 21 & \cite{choi2024iterative,   hemberg2024evolving, han2024generating, van2024llamea, jain2022jigsaw,   palkowski2024gpt, sun2024autosat, xu2024large, qu2024dynamic, li2024falcon,   DBLP:journals/corr/abs-2406-07496, DBLP:journals/corr/abs-2405-15189,   DBLP:conf/nips/MadaanTGHGW0DPY23, DBLP:conf/nips/ShinnCGNY23,   DBLP:conf/apsec/ShirafujiOSMW23,   DBLP:journals/nature/RomeraParedesBNBKDREWFKF24,   DBLP:journals/corr/abs-2401-08500, DBLP:journals/corr/abs-2406-12502,   zhang2024revolve, van2024loop, peng2024perfcodegen} \\ \cline{3-5} 
 &  & Compiler & 4 & \cite{DBLP:conf/hpec/GuoM22,   DBLP:journals/corr/abs-2309-07062, DBLP:journals/corr/abs-2403-14714,   DBLP:journals/corr/abs-2407-02524} \\ \cline{3-5} 
 &  & Decoder & 1 & \cite{pan2024ecodemasteringefficientcode} \\ \cline{3-5} 
 &  & Diff   generator & 1 & \cite{DBLP:journals/corr/abs-2309-14846} \\ \hline
Evaluation & 10 & Evaluator & 10 & \cite{hemberg2024evolving,   sun2024autosat, xu2024code, peng2024large,   pan2024ecodemasteringefficientcode, DBLP:journals/corr/abs-2406-07496,   DBLP:conf/nips/MadaanTGHGW0DPY23, DBLP:conf/nips/ShinnCGNY23,   DBLP:journals/corr/abs-2406-04693, zhang2024revolve} \\ \hline
\multirow{3}{*}{Preprocessing} & \multirow{3}{*}{6} & Advisor & 2 & \cite{sun2024autosat,   DBLP:journals/corr/abs-2406-04693} \\ \cline{3-5} 
 &  & Encoder & 2 & \cite{xu2024optimizing,   pan2024ecodemasteringefficientcode} \\ \cline{3-5} 
 &  & Type   inferencer & 2 & \cite{DBLP:journals/pacmpl/YeZSS23,   DBLP:journals/concurrency/HanLDLZ24}
\\
\bottomrule
\end{tabular}
\end{adjustbox}
\label{tb:role_LMs}
\end{table}

\subsubsection{Preprocessing}
LMs also played key roles in the preprocessing category, as shown in six studies. In this category, LMs handle roles including: \textbf{advisor}, which is often used in an agentic flow and provides guidance and oversight to the other LM agents~\cite{sun2024autosat, DBLP:journals/corr/abs-2406-04693}; \textbf{encoder}, which extracts hidden representations from different code contexts to help understand the semantics of the code~\cite{xu2024optimizing,   pan2024ecodemasteringefficientcode}; and \textbf{type inferencer}, which leverages patterns in variable names and code structures to infer the concrete variable types without requiring explicit type annotations from programmers~\cite{DBLP:journals/pacmpl/YeZSS23,   DBLP:journals/concurrency/HanLDLZ24}. These preprocessing activities provide the necessary groundwork for tools to operate effectively in subsequent tasks like decoding, evaluation, or optimization~\cite{pan2024ecodemasteringefficientcode}. 

\keybox{
\faIcon{search} \textit{\textbf{Finding \thefindingcount:} Generation-related LMs played the most fundamental roles (73 instances) and are often combined with various assist tools. Evaluator LMs can be insightful and explainable but may face potential accuracy issues (10 studies). LM-based preprocessing offers foundations for subsequent tasks, but demands significant computational power (six studies).}
\addtocounter{findingcount}{1}
}

\suggestbox{
\faIcon{thumbs-up} \textit{\textbf{Recommendation \thesuggestioncount:} Future research can combine the strengths of different LM roles to address complex code optimization issues more effectively, and leverage various tools to mitigate the drawbacks of each role.}
\addtocounter{suggestioncount}{1}
}

\section{RQ3: How Was the Code Optimization Problem Defined?}
\label{sec:RQ3_problem_definition}
In this section, we investigate the programming languages and performance metrics involved when applying LMs to code optimization. Understanding these questions is critical as it helps researchers and practitioners select appropriate strategies and tailor their techniques to specific settings, ultimately advancing the field of code performance optimization.

\subsection{RQ3.1: What Programming Languages Were Considered?}
\label{subsec:programming_languages}
We first list the programming languages optimized using LMs in Table~\ref{tb:languages}, and then in Figure~\ref{fig:num_languages}, we present the number of languages that are involved in each primary study. 

\subsubsection{Targeted programming languages}
From Table~\ref{tb:languages}, it is evident that most studies focused on \textbf{high-level languages}, accounting for 53 instances in total. Within this category, Python dominated with 30 studies, due to its wide use in data science, machine learning, and scripting, making it an ideal candidate for optimization tasks~\cite{hemberg2024evolving, xu2024large, qu2024dynamic, li2024falcon}. C++ and C, being foundational systems programming languages, were targeted in nine and six studies respectively due to their heavy usage in performance-critical applications~\cite{peng2024large, DBLP:journals/corr/abs-2309-14846, DBLP:conf/hpec/GuoM22}, while Rust, C\#, and Java represent emerging or enterprise-level programming tools that also require optimization~\cite{DBLP:conf/nips/ShinnCGNY23, DBLP:journals/corr/abs-2306-17077, DBLP:journals/corr/abs-2404-18864}. 

In contrast, \textbf{low-level languages} were addressed in only six studies, with four focusing on LLVM-IR, which is an intermediate representation that abstracts a simplified view of programs, enabling compiler-level optimization~\cite{DBLP:conf/euromlsys/GrubisicSSLMC24, DBLP:journals/corr/abs-2309-07062, DBLP:journals/corr/abs-2403-14714, DBLP:journals/corr/abs-2407-02524}; and two focusing on assembly code, which uses mnemonic codes, labels, and directives to represent instructions and data structures, allowing for fine-tuned control over hardware~\cite{DBLP:journals/corr/abs-2109-13498, DBLP:journals/corr/abs-2407-02524}. This focus on high-level languages offers advantages such as a larger user base and more available benchmarks for evaluation. However, it may also result in an underrepresentation of low-level optimization challenges, limiting innovations for languages closer to hardware~\cite{DBLP:conf/hpec/GuoM22, DBLP:journals/corr/abs-2407-02524, wei2024improving}.

Additionally, \textbf{domain-specific languages (DSLs)} were considered in another six studies, each targeting a unique DSL such as tensor processing code, mapper code, heuristic code et al.~\cite{DBLP:journals/corr/abs-2408-03408, wei2024improving, sun2024autosat, xu2024optimizing, yao2024rtlrewritermethodologieslargemodels, DBLP:journals/concurrency/HanLDLZ24}. The diversity of DSLs shows that LM-based code optimization is beneficial in different domains, yet, since each study is typically constrained to a narrow scope, the broader applicability of these techniques may be limited.

\begin{table}[t!]
  \caption{Distribution of optimized languages (one study might be in multiple categories).}
\centering
\footnotesize
\begin{adjustbox}{width=\linewidth,center}
\begin{tabular}{lllll}
\toprule
\textbf{Category} & \textbf{Total \#} & \textbf{Languague} & \textbf{\# studies} & \textbf{Reference} \\ \hline
\multirow{6}{*}{High-level languages} & \multirow{6}{*}{53} & Python & 30 & \cite{DBLP:journals/corr/abs-2310-02304,   pan2024measuringcodeefficiencyoptimization,   pan2024ecodemasteringefficientcode,   shivashankar2024betterpythonprogrammingall,   DBLP:journals/corr/abs-2406-07496, gao2024searchbasedllmscodeoptimization,   DBLP:journals/corr/abs-2405-15189, DBLP:conf/nips/MadaanTGHGW0DPY23,   li2024instructcoderinstructiontuninglarge, DBLP:conf/nips/ShinnCGNY23,   DBLP:conf/apsec/ShirafujiOSMW23,   DBLP:journals/nature/RomeraParedesBNBKDREWFKF24,   DBLP:journals/corr/abs-2404-18864, DBLP:journals/corr/abs-2312-05657,   ishida2024langpropcodeoptimizationframework,   DBLP:journals/corr/abs-2407-14044, DBLP:journals/pacmpl/YeZSS23,   DBLP:journals/corr/abs-2406-12502, DBLP:journals/corr/abs-2401-08500,   hemberg2024evolving, han2024generating, van2024llamea, jain2022jigsaw,   xu2024large, qu2024dynamic, li2024falcon, huang2024effi, zhang2024codedpo,   van2024loop, peng2024perfcodegen} \\ \cline{3-5} 
 &  & C++ & 9 & \cite{DBLP:journals/corr/abs-2309-14846,   gao2024searchbasedllmscodeoptimization, DBLP:journals/corr/abs-2404-18864,   DBLP:conf/iclr/ShypulaMZ0GYHNR24, DBLP:journals/corr/abs-2406-11935,   DBLP:journals/corr/abs-2401-08500, xu2024code, palkowski2024gpt, peng2024large} \\ \cline{3-5} 
 &  & C & 6 & \cite{DBLP:journals/corr/abs-2309-14846,   DBLP:conf/hpec/GuoM22, DBLP:journals/corr/abs-2406-04693,   DBLP:journals/corr/abs-2401-08500, DBLP:journals/concurrency/HanLDLZ24,   palkowski2024gpt} \\ \cline{3-5} 
 &  & Rust & 3 & \cite{DBLP:journals/corr/abs-2406-07496,   DBLP:conf/nips/ShinnCGNY23, DBLP:journals/corr/abs-2401-08500} \\ \cline{3-5} 
 &  & C\# & 3 & \cite{DBLP:conf/sigsoft/GargMCSW22,   DBLP:journals/corr/abs-2401-08500, DBLP:journals/corr/abs-2306-17077} \\ \cline{3-5} 
 &  & Java & 2 & \cite{DBLP:journals/corr/abs-2404-18864,   choi2024iterative} \\ \hline
\multirow{2}{*}{Low-level   languages} & \multirow{2}{*}{6} & LLVM-IR & 4 & \cite{DBLP:conf/euromlsys/GrubisicSSLMC24,   DBLP:journals/corr/abs-2309-07062, DBLP:journals/corr/abs-2403-14714,   DBLP:journals/corr/abs-2407-02524} \\ \cline{3-5} 
 &  & Assembly   code & 2 & \cite{DBLP:journals/corr/abs-2109-13498,   DBLP:journals/corr/abs-2407-02524} \\ \hline
\multirow{6}{*}{Domain-specific languages} & \multirow{6}{*}{6} & Tensor processing code & 1 & \cite{DBLP:journals/corr/abs-2408-03408} \\ \cline{3-5} 
 &  & Mapper   code & 1 & \cite{wei2024improving} \\ \cline{3-5} 
 &  & Heuristsic   code & 1 & \cite{sun2024autosat} \\ \cline{3-5} 
 &  & High-Level   Synthesis (HSL) & 1 & \cite{xu2024optimizing} \\ \cline{3-5} 
 &  & Register   Transfer Level (RTL) & 1 & \cite{yao2024rtlrewritermethodologieslargemodels} \\ \cline{3-5} 
 &  & Structured   Text (ST) & 1 & \cite{DBLP:journals/concurrency/HanLDLZ24}
\\
\bottomrule
\end{tabular}
\end{adjustbox}
\label{tb:languages}
\end{table}

\subsubsection{The number of languages}
Except for the specific languages being optimized, it is also valuable to investigate the number of languages optimized per study. As shown in Figure~\ref{fig:num_languages}, most studies (81\%) focused on a \textbf{single language}, reflecting the difficulty in generalizing optimization techniques across multiple languages with different syntaxes, semantics, and performance characteristics~\cite{pan2024ecodemasteringefficientcode, shivashankar2024betterpythonprogrammingall, DBLP:journals/corr/abs-2109-13498, DBLP:conf/hpec/GuoM22, DBLP:journals/corr/abs-2405-15189}. Seven studies (13\%) targeted \textbf{two languages}, often pairing languages with complementary use cases or interoperable ecosystems, such as Python and C++~\cite{gao2024searchbasedllmscodeoptimization}, C and C++~\cite{palkowski2024gpt, DBLP:journals/corr/abs-2309-14846}, Python and Rust~\cite{DBLP:conf/nips/ShinnCGNY23, DBLP:journals/corr/abs-2406-07496}, ST and C~\cite{DBLP:journals/concurrency/HanLDLZ24}, or IR and assembly code~\cite{DBLP:journals/corr/abs-2407-02524}. In contrast, only one study handled \textbf{three languages}~\cite{DBLP:journals/corr/abs-2404-18864}, and two studies remained \textbf{unclear} in their focuses~\cite{DBLP:journals/corr/abs-2401-08500, zhang2024revolve}. These multi-lingual techniques, while rare, offer the potential for broader applicability, but they may face challenges in balancing robustness and accuracy across languages.

\keybox{
\faIcon{search} \textit{\textbf{Finding \thefindingcount:} (1) LM-based code optimization primarily targeted high-level languages (53 instances) due to their widespread usage and accessible datasets. (2) The prevalence of single-language studies (81\%) highlights the challenges of achieving generalizability across languages.}
\addtocounter{findingcount}{1}
}

\suggestbox{
\faIcon{thumbs-up} \textit{\textbf{Recommendation \thesuggestioncount:} (1) Low-level languages and DSLs, while less represented (six studies each), often address critical optimization needs in their specific areas, requiring future attention. (2) The limited number of multi-language studies suggests the potential for developing cross-language code optimization frameworks.}
\addtocounter{suggestioncount}{1}
}

\begin{figure}[!t]
\centering
\begin{minipage}{0.4\textwidth}
\begin{tikzpicture}[scale=0.8]
\tikzstyle{every node}=[font=\large]
\definecolor{beaublue}{rgb}{0.74, 0.83, 0.9}
\definecolor{applegreen}{rgb}{0.55, 0.71, 0.0}
\definecolor{dollarbill}{rgb}{0.52, 0.73, 0.4}
\definecolor{ao(english)}{rgb}{0.0, 0.5, 0.0}
\definecolor{amaranth}{rgb}{0.9, 0.17, 0.31}
\definecolor{cerublue}{rgb}{0.16, 0.32, 0.75}
\definecolor{frenchblue}{rgb}{0.0, 0.45, 0.73}
\definecolor{iceberg}{rgb}{0.44, 0.65, 0.82}
  
  
    \pie[
        /tikz/every pin/.style={align=center},
        text=pin,
        rotate=20,
        explode=0, 
        color={cerublue!70, iceberg!50},
        ]
        {
        81/\texttt{One} (43),
        13/\texttt{Two} (7)
        }

    \pie[
        /tikz/every pin/.style={align=center},
        text=pin,
        before number=\phantom,
        after number=,
        rotate=358.4, 
        explode=0, 
        color={frenchblue!50, red!60, red!40, orange!50, yellow!50, lime!50, green!50, teal!50, cyan!50, purple!50, magenta!50, pink!50},
        ]
        {
        2/\texttt{2\% Three} (1),
        4/\texttt{4\% Unclear} (2)
        }

    \end{tikzpicture}
  \caption{Distribution of \# optimized programming languages.}
 \label{fig:num_languages}
  \end{minipage}
~\hspace{0.5cm}
\begin{minipage}{0.36\textwidth}
\begin{tikzpicture}[scale=0.8]
\tikzstyle{every node}=[font=\large]
\definecolor{beaublue}{rgb}{0.74, 0.83, 0.9}
\definecolor{applegreen}{rgb}{0.55, 0.71, 0.0}
\definecolor{dollarbill}{rgb}{0.52, 0.73, 0.4}
\definecolor{ao(english)}{rgb}{0.0, 0.5, 0.0}
\definecolor{amaranth}{rgb}{0.9, 0.17, 0.31}
\definecolor{cerublue}{rgb}{0.16, 0.32, 0.75}
\definecolor{frenchblue}{rgb}{0.0, 0.45, 0.73}
\definecolor{iceberg}{rgb}{0.44, 0.65, 0.82}
  
  
    \pie[
        /tikz/every pin/.style={align=center},
        text=pin,
        rotate=30,
        explode=0, 
        color={cerublue!70, iceberg!50, frenchblue!50, red!60},
        ]
        {
        79/\texttt{One} (42),
        17/\texttt{Two} (9),
        4/\texttt{Three} (2)
        }


    \end{tikzpicture}
\vspace{-0.4cm}
  \caption{Distribution of \# targeted performance metrics.}
 \label{fig:num_performance_metrics}
  \end{minipage}
 \end{figure}

\subsection{RQ3.2: What Performance Metrics Were Optimized?}
\label{subsec:performance_metrics}
Performance metrics serve as the foundation for assessing the effectiveness of optimization techniques, hence, we survey the performance metrics targeted for optimization and the number of optimized metrics in each primary study, as shown in Table~\ref{tb:performance_metrics} and Figure~\ref{fig:num_performance_metrics}. 

\subsubsection{The optimized performance metrics}
Among others, \textbf{efficiency-related metrics} were the most commonly explored, addressed 27 times. Within this category, runtime was the dominant metric, used in 24 studies, due to its direct impact on user experience and the widespread relevance of reducing execution time in various applications~\cite{DBLP:conf/nips/MadaanTGHGW0DPY23, DBLP:journals/corr/abs-2401-08500, DBLP:journals/corr/abs-2404-18864, DBLP:journals/corr/abs-2406-04693, DBLP:journals/corr/abs-2312-05657}. Latency and throughput, which are often critical in real-time applications or high-performance computing, were used in two and one studies respectively~\cite{peng2024large, xu2024optimizing, wei2024improving}. These results reflect the importance of runtime, but they also indicate a potential overlook of other critical efficiency metrics.

\textbf{General quality metrics} were addressed 16 times, being the second most explored category. Specifically, five studies each opted to optimize code size, which is often measured by the count of instructions or binary size~\cite{DBLP:journals/corr/abs-2403-14714, DBLP:journals/corr/abs-2406-12502, DBLP:journals/corr/abs-2407-02524}, and cyclomatic complexity, which measures the number of linearly independent paths in the code~\cite{shivashankar2024betterpythonprogrammingall, DBLP:conf/apsec/ShirafujiOSMW23, choi2024iterative}. Readability and maintainability, while less frequently mentioned in three studies each, are essential for avoiding technical debt, i.e., the future cost of fixing code due to choosing easy but suboptimal solutions in the past~\cite{shivashankar2024betterpythonprogrammingall, DBLP:journals/concurrency/HanLDLZ24, li2024falcon}. The results suggest an awareness of the importance of long-term usability and developer experience, however, these metrics are more subjective and harder to quantify compared to efficiency.

Moreover, \textbf{task-specific metrics} were explored in 14 studies, covering a wide range of specialized objectives achieved by the optimized code, such as task completion rate~\cite{xu2024large, zhang2024revolve}, convergence quality~\cite{van2024llamea, van2024loop}, synthesis accuracy~\cite{jain2022jigsaw}, and success rate~\cite{han2024generating}, reflecting the diversity and specificity of tasks targeted by LM-based code optimization. For example,~\citet{DBLP:journals/corr/abs-2408-03408} aimed to optimize the performance of DSLs for hardware accelerators by leveraging feedback from a hardware cost model called Ansor~\cite{zheng2020ansor}. Nonetheless, while these domain-specific metrics allow for precise evaluation in specific contexts, their narrow scope can hinder the generalization of the evaluation results and findings.

Lastly, \textbf{resource usage metrics} were addressed in nine studies, where memory usage was the most studied (six studies), due to its direct impact on the scalability and practicality of programs, particularly in resource-constrained environments~\cite{DBLP:conf/sigsoft/GargMCSW22, huang2024effi} and some competitive programming tasks~\cite{DBLP:journals/corr/abs-2309-14846, DBLP:journals/corr/abs-2405-15189, DBLP:journals/corr/abs-2401-08500, DBLP:journals/corr/abs-2407-14044}. CPU usage and energy consumption, while less frequently examined (two and one studies respectively), are critical in contexts where hardware efficiency or sustainability is a priority, such as mobile, embedded, or high-performance computing systems~\cite{DBLP:journals/corr/abs-2306-17077, DBLP:conf/sigsoft/GargMCSW22, peng2024large}. Noteworthy, these results emphasize a lack of attention to the code's sustainability and resource efficiency, which can be challenging to profile with specialized tools.

\begin{table}[t!]
  \caption{Distribution of targeted performance metrics (one study might be in multiple categories).}
\centering
\footnotesize
\begin{adjustbox}{width=\linewidth,center}
\begin{tabular}{lllll}
\toprule
\textbf{Category} & \textbf{Total \#} & \textbf{Metric} & \textbf{\# studies} & \textbf{Reference} \\ \hline
\multirow{3}{*}{Efficiency} & \multirow{3}{*}{27} & Runtime & 24 & \cite{pan2024measuringcodeefficiencyoptimization,   DBLP:journals/corr/abs-2406-07496, DBLP:journals/corr/abs-2309-14846,   gao2024searchbasedllmscodeoptimization, DBLP:journals/corr/abs-2109-13498,   DBLP:journals/corr/abs-2405-15189, DBLP:conf/nips/MadaanTGHGW0DPY23,   DBLP:journals/corr/abs-2401-08500, DBLP:journals/corr/abs-2404-18864,   DBLP:journals/corr/abs-2406-04693, DBLP:journals/corr/abs-2312-05657,   DBLP:conf/iclr/ShypulaMZ0GYHNR24, DBLP:journals/corr/abs-2406-11935,   DBLP:journals/corr/abs-2407-14044, DBLP:journals/corr/abs-2406-12502,   hemberg2024evolving, palkowski2024gpt, xu2024code, qu2024dynamic,   huang2024effi, zhang2024codedpo, DBLP:journals/corr/abs-2310-02304,   pan2024ecodemasteringefficientcode, peng2024perfcodegen} \\ \cline{3-5} 
 &  & Latency & 2 & \cite{peng2024large,   xu2024optimizing} \\ \cline{3-5} 
 &  & Throughput & 1 & \cite{wei2024improving} \\ \hline
\multirow{4}{*}{General quality} & \multirow{4}{*}{16} & Code size & 5 & \cite{DBLP:conf/euromlsys/GrubisicSSLMC24,   DBLP:journals/corr/abs-2309-07062, DBLP:journals/corr/abs-2403-14714,   DBLP:journals/corr/abs-2406-12502, DBLP:journals/corr/abs-2407-02524} \\ \cline{3-5} 
 &  & Complexity & 5 & \cite{shivashankar2024betterpythonprogrammingall,   DBLP:conf/apsec/ShirafujiOSMW23, choi2024iterative, li2024falcon,   DBLP:journals/nature/RomeraParedesBNBKDREWFKF24} \\ \cline{3-5} 
 &  & Readability & 3 & \cite{shivashankar2024betterpythonprogrammingall,   DBLP:journals/concurrency/HanLDLZ24, li2024falcon} \\ \cline{3-5} 
 &  & Maintainability & 3 & \cite{shivashankar2024betterpythonprogrammingall,   DBLP:journals/concurrency/HanLDLZ24, li2024falcon} \\ \hline
\multirow{12}{*}{Task-specific} & \multirow{12}{*}{14} & Task completion rate & 2 & \cite{xu2024large,   zhang2024revolve} \\ \cline{3-5} 
 &  & Convergence   quality & 2 & \cite{van2024llamea,   van2024loop} \\ \cline{3-5} 
 &  & Synthesis   accuracy & 1 & \cite{jain2022jigsaw} \\ \cline{3-5} 
 &  & Number   of instances solved & 1 & \cite{sun2024autosat} \\ \cline{3-5} 
 &  & Success   rate & 1 & \cite{han2024generating} \\ \cline{3-5} 
 &  & Synthesis   performance & 1 & \cite{yao2024rtlrewritermethodologieslargemodels} \\ \cline{3-5} 
 &  & Hardware   performance & 1 & \cite{DBLP:journals/corr/abs-2408-03408} \\ \cline{3-5} 
 &  & Reference   match & 1 & \cite{DBLP:conf/hpec/GuoM22} \\ \cline{3-5} 
 &  & Code   edit accuracy & 1 & \cite{li2024instructcoderinstructiontuninglarge} \\ \cline{3-5} 
 &  & Decision-making   performance & 1 & \cite{DBLP:conf/nips/ShinnCGNY23} \\ \cline{3-5} 
 &  & Driving   score & 1 & \cite{ishida2024langpropcodeoptimizationframework} \\ \cline{3-5} 
 &  & Type   inference speed & 1 & \cite{DBLP:journals/pacmpl/YeZSS23} \\ \hline
\multirow{3}{*}{Resource usage} & \multirow{3}{*}{9} & Memory usage & 6 & \cite{DBLP:journals/corr/abs-2309-14846,   DBLP:journals/corr/abs-2405-15189, DBLP:journals/corr/abs-2401-08500,   DBLP:journals/corr/abs-2407-14044, DBLP:conf/sigsoft/GargMCSW22,   huang2024effi} \\ \cline{3-5} 
 &  & CPU   usage & 2 & \cite{DBLP:journals/corr/abs-2306-17077,   DBLP:conf/sigsoft/GargMCSW22} \\ \cline{3-5} 
 &  & Energy & 1 & \cite{peng2024large}
\\
\bottomrule
\end{tabular}
\end{adjustbox}
\label{tb:performance_metrics}
\end{table}

\subsubsection{The number of optimized metrics}
Figure~\ref{fig:num_performance_metrics} provides additional insights into the number of targeted performance metrics across studies. Notably, most studies (42, or 79\%) focused on a \textbf{single performance metric}~\cite{DBLP:conf/nips/MadaanTGHGW0DPY23, DBLP:conf/apsec/ShirafujiOSMW23, DBLP:conf/euromlsys/GrubisicSSLMC24, DBLP:journals/corr/abs-2306-17077, DBLP:journals/corr/abs-2404-18864}, nine studies (17\%) explored \textbf{two metrics}~\cite{peng2024large, huang2024effi, DBLP:journals/corr/abs-2309-14846, DBLP:journals/corr/abs-2405-15189}, while only two studies (4\%) addressed \textbf{three metrics}~\cite{li2024falcon, shivashankar2024betterpythonprogrammingall}. The predominance of single-metric studies reflects a focus on specific goals, yet it may limit the ability to capture holistic improvements, as optimizing one metric (e.g., runtime) may negatively impact others (e.g., CPU usage)~\cite{peng2024large}. Hence, it is essential to balance multiple metrics and resolve conflicts between competing objectives.

\keybox{
\faIcon{search} \textit{\textbf{Finding \thefindingcount:} The results reveal a strong emphasis on single performance metrics (79\%), mainly code efficiency-related ones (27 instances), which reflects the importance of runtime performance in most optimization tasks and the conflicting nature of different metrics.}
\addtocounter{findingcount}{1}
}

\suggestbox{
\faIcon{thumbs-up} \textit{\textbf{Recommendation \thesuggestioncount:} The limited focus on multi-metric optimization suggests an opportunity for future research to develop balanced techniques that account for diverse performance objectives.}
\addtocounter{suggestioncount}{1}
}




\section{RQ4: How Were the Proposed Code Optimization Methods Evaluated?}
\label{sec:RQ4_evaluation_method}
The evaluation methodology is also vital as it ensures the credibility, practical applicability, strengths, and weaknesses of the proposed techniques, guiding future enhancements in both research and practical applications. Therefore, this section examines the datasets/benchmarks used for evaluation, whether they are evaluated in real-world data, and the performance metrics employed.

\begin{table}[t!]
  \caption{Distribution of datasets and benchmarks (one study might be in multiple categories).}
\centering
\footnotesize
\begin{adjustbox}{width=\linewidth,center}
\begin{tabular}{p{1.6cm}llllllll}
\toprule
\textbf{Category} & \textbf{Total \#} & \textbf{Dataset} & \textbf{Source} & \textbf{Size} & \textbf{Languages} & \textbf{Performance} & \textbf{Repo} & \textbf{Reference} \\ \hline
\multirow{16}{1.6cm}{Competitive programming} & \multirow{16}{*}{35} & HumanEval~\cite{DBLP:journals/corr/abs-2107-03374} & Hand-crafted by experts & 164 programming tasks & Python & Correctness & \href{https://github.com/openai/human-eval}{Link} & \cite{DBLP:journals/corr/abs-2405-15189,   DBLP:conf/nips/ShinnCGNY23, DBLP:journals/corr/abs-2406-12502, li2024falcon,   huang2024effi, zhang2024codedpo, peng2024perfcodegen} \\ \cline{3-9} 
 &  & MBPP~\cite{DBLP:journals/corr/abs-2108-07732} & Programming problems & 974 programming tasks & Python & Correctness & \href{https://github.com/google-research/google-research/tree/master/mbpp}{Link} & \cite{DBLP:journals/corr/abs-2405-15189,   DBLP:conf/nips/ShinnCGNY23, DBLP:journals/corr/abs-2406-12502, li2024falcon,   zhang2024codedpo, peng2024perfcodegen} \\ \cline{3-9} 
 &  & PIE~\cite{DBLP:conf/iclr/ShypulaMZ0GYHNR24} & CodeNet & 77K pairs of slow-fast code & C++ & Runtime & \href{https://github.com/LearningOpt/pie?tab=readme-ov-file#dataset}{Link} & \cite{gao2024searchbasedllmscodeoptimization,   DBLP:conf/nips/MadaanTGHGW0DPY23, DBLP:journals/corr/abs-2312-05657,   DBLP:conf/iclr/ShypulaMZ0GYHNR24} \\ \cline{3-9} 
 &  & LeetcodeHardGym~\cite{DBLP:conf/nips/ShinnCGNY23} & Leetcode & 40 questions & Python, Rust & Runtime & \href{https://github.com/GammaTauAI/leetcode-hard-gym}{Link} & \cite{DBLP:conf/nips/ShinnCGNY23,   DBLP:journals/corr/abs-2406-07496, zhang2024revolve} \\ \cline{3-9} 
 &  & EffiBench~\cite{DBLP:journals/corr/abs-2402-02037} & Leetcode & 1K efficiency-critical coding   problems & Python & Runtime, memory & \href{https://huggingface.co/datasets/DONG19/EffiBench}{Link} & \cite{DBLP:journals/corr/abs-2405-15189,   huang2024effi} \\ \cline{3-9} 
 &  & CodeContests~\cite{DBLP:journals/corr/abs-2203-07814} & Aizu Online Judge, AtCoder & 13,610 samples & Python, C++, Java & Runtime, memory & \href{https://github.com/google-deepmind/code_contests}{Link} & \cite{DBLP:conf/apsec/ShirafujiOSMW23,   DBLP:journals/corr/abs-2401-08500} \\ \cline{3-9} 
 &  & APPS~\cite{hendrycks2021measuring} & Coding websites & 10k coding problems & Python & Correctness & \href{https://github.com/hendrycks/apps?form=MG0AV3}{Link} & \cite{li2024falcon,   peng2024perfcodegen} \\ \cline{3-9} 
 &  & ECCO~\cite{DBLP:journals/corr/abs-2407-14044} & CodeNet & 50K solution pairs & Python & Runtime, memory & \href{https://github.com/CodeEff/ECCO}{Link} & \cite{DBLP:journals/corr/abs-2407-14044} \\ \cline{3-9} 
 &  & FunSearch~\cite{DBLP:journals/nature/RomeraParedesBNBKDREWFKF24} & Algorithmic problems & $10^6$ samples & Python & Complexity, readability,   maintainability & \href{https://github.com/google-deepmind/funsearch}{Link} & \cite{DBLP:journals/nature/RomeraParedesBNBKDREWFKF24} \\ \cline{3-9} 
 &  & Supersonic~\cite{DBLP:journals/corr/abs-2309-14846} & CodeNet & 314,435 samples & C, C++ & Runtime, memory & \href{https://github.com/ASSERT-KTH/Supersonic}{Link} & \cite{DBLP:journals/corr/abs-2309-14846} \\ \cline{3-9} 
 &  & GEC~\cite{DBLP:conf/internetware/PanL23} & CodeForces & 31,577 pairs of slow-fast   code & Python & Runtime & \href{https://github.com/CodeGeneration2/GEC-Dataset}{Link} & \cite{pan2024ecodemasteringefficientcode} \\ \cline{3-9} 
 &  & CodeNet~\cite{DBLP:journals/corr/abs-2105-12655} & AIZU Online Judge, AtCoder & 14 million samples & C++, C, C\#, Python, Java... & Runtime, memory, code size & \href{https://developer.ibm.com/exchanges/data/all/project-codenet/}{Link} & \cite{DBLP:conf/hpec/GuoM22} \\ \cline{3-9} 
 &  & ACEOB~\cite{pan2024measuringcodeefficiencyoptimization} & CodeForces & 95,359 pairs of   efficient-inefficient code & Python & Runtime & \xmark & \cite{pan2024measuringcodeefficiencyoptimization} \\ \cline{3-9} 
 &  & Effi-Code~\cite{huang2024effi} & Coding datasets & 9,451 tasks & Python & Runtime, memory & \href{https://github.com/huangd1999/Effi-Code}{Link} & \cite{huang2024effi} \\ \cline{3-9} 
 &  & SAPIE~\cite{xu2024code} & CodeNet & 77k pairs of slow-fast code & C++ & Runtime & \xmark & \cite{xu2024code} \\ \cline{3-9} 
 &  & PIE-problem~\cite{DBLP:journals/corr/abs-2406-11935} & CodeNet & 18,242 pairs of slow-fast   code & C++ & Runtime & \xmark & \cite{DBLP:journals/corr/abs-2406-11935} \\ \hline
\multirow{12}{*}{General SE} & \multirow{12}{*}{13} & DeepDev-PERF~\cite{DBLP:conf/sigsoft/GargMCSW22} & GitHub & 45k open-source repositories & C\# & CPU, memory & \xmark & \cite{DBLP:journals/corr/abs-2306-17077,   DBLP:conf/sigsoft/GargMCSW22} \\ \cline{3-9} 
 &  & AnghaBench~\cite{DBLP:conf/cgo/SilvaKMRGP21} & GitHub & 1 million samples & C & Runtime, code size & \href{http://cuda.dcc.ufmg.br/angha/benchmarks}{Link} & \cite{DBLP:conf/hpec/GuoM22} \\ \cline{3-9} 
 &  & InstructCoder~\cite{li2024instructcoderinstructiontuninglarge} & GitHub & 114K instruction-input-output   triplets & Python & Complexity, readability,   maintainability & \href{https://github.com/qishenghu/InstructCoder}{Link} & \cite{li2024instructcoderinstructiontuninglarge} \\ \cline{3-9} 
 &  & Energy-Language~\cite{pereira2017energy} & Software repositories & 10 problems & 27 languages & Energy, memory, runtime & \href{https://github.com/greensoftwarelab/Energy-Languages?tab=readme-ov-file#checking-energy-consumption-in-programming-languages-using-the-computer-language-benchmark-game-as-a-case-study}{Link} & \cite{peng2024large} \\ \cline{3-9} 
 &  & BetterPython~\cite{shivashankar2024betterpythonprogrammingall} & CommitPackFT, CodeAlpaca & 34,139 samples & Python & Complexity, readability,   maintainability & \href{https://zenodo.org/records/10940620}{Link} & \cite{shivashankar2024betterpythonprogrammingall} \\ \cline{3-9} 
 &  & Defects4J~\cite{just2014defects4j} & Open-srouce projects & 17 projects & Java & Complexity & \href{https://github.com/rjust/defects4j}{Link} & \cite{choi2024iterative} \\ \cline{3-9} 
 &  & PP4F~\cite{kastner2018parallel} & Synthesis & 699 examples & HLS & Latency & \href{https://github.com/KastnerRG/pp4fpgas?form=MG0AV3}{Link} & \cite{xu2024optimizing} \\ \cline{3-9} 
 &  & RewriterBench~\cite{yao2024rtlrewritermethodologieslargemodels} & Industry cases & 55 cases & RTL & Synthesis performance & \href{https://github.com/yaoxufeng/RTLRewriter-Bench}{Link} & \cite{yao2024rtlrewritermethodologieslargemodels} \\ \cline{3-9} 
 &  & ST-to-C~\cite{DBLP:journals/concurrency/HanLDLZ24} & Industry cases & 3 case studies & Structured Text (ST), C & Readability, maintainability & \xmark & \cite{DBLP:journals/concurrency/HanLDLZ24} \\ \cline{3-9} 
 &  & PandasEval~\cite{jain2022jigsaw} & StackOverflow, Hackathon & 89 Pandas tasks & Python & Correctness & \href{https://github.com/microsoft/JigsawDataset}{Link} & \cite{jain2022jigsaw} \\ \cline{3-9} 
 &  & Big   Assembly~\cite{DBLP:journals/corr/abs-2109-13498} & GitHub & 25,141 assembly functions & x86-64 assembly language & CPU-clock cycles & \xmark & \cite{DBLP:journals/corr/abs-2109-13498} \\ \cline{3-9} 
 &  & CSmith~\cite{DBLP:conf/pldi/YangCER11} & Synthesis & Unlimited & C & Runtime & \href{https://github.com/csmith-project/csmith}{Link} & \cite{DBLP:conf/hpec/GuoM22} \\ \hline
\multirow{4}{*}{Compiler} & \multirow{4}{*}{7} & PolyBench~\cite{DBLP:conf/cc/Abella-Gonzalez21} & Synthesis & 30 numerical polyhedral kernels & Python, C & Runtime, memory & \href{https://github.com/UDC-GAC/polybench-python}{Link} & \cite{DBLP:journals/corr/abs-2408-03408,   DBLP:journals/corr/abs-2404-18864, DBLP:journals/pacmpl/YeZSS23} \\ \cline{3-9} 
 &  & LLM4Compiler~\cite{DBLP:journals/corr/abs-2309-07062} & GitHub, synthesis & 1 million  functions & LLVM-IR & Code size & \xmark & \cite{DBLP:journals/corr/abs-2309-07062,   DBLP:journals/corr/abs-2403-14714} \\ \cline{3-9} 
 &  & TSVC~\cite{maleki2011evaluation} & Synthesis & 149 test cases & C & Runtime, code size & \href{https://github.com/UoB-HPC/TSVC_2}{Link} & \cite{DBLP:journals/corr/abs-2406-04693} \\ \cline{3-9} 
 &  & Priority   Sampling~\cite{DBLP:conf/euromlsys/GrubisicSSLMC24} & GitHub & 50K functions & LLVM-IR & Code size & \xmark & \cite{DBLP:conf/euromlsys/GrubisicSSLMC24} \\ \hline
\multirow{2}{*}{Data science} & \multirow{2}{*}{2} & Big-DS-1000~\cite{qu2024dynamic} & StackOverflow & 1000 data science problems & Python & Runtime & \xmark & \cite{qu2024dynamic} \\ \cline{3-9} 
 &  & DS-1000~\cite{lai2023ds} & StackOverflow & 1000 data science problems & Python & Correctness & \href{https://ds1000-code-gen.github.io/}{Link} & \cite{zhang2024codedpo}
\\
\bottomrule
\end{tabular}
\end{adjustbox}
\label{tb:datasets}
\end{table}

\subsection{RQ4.1: What Were the Existing Datasets and Benchmarks?}
\label{subsec:datasets}
In this subsection, we identify and categorize the datasets/benchmarks used according to their application domains. To provide a comprehensive overview of their advantages and characteristics, we also collected detailed information such as the original proposed papers, code sources, dataset sizes, programming languages, performance data, and repository links, as shown in Table~\ref{tb:datasets}.

\subsubsection{Competitive programming datasets}
Datasets for competitive programming tasks were the most commonly used, which consist of a set of problem statements, testing cases, source code, and performance data, with 35 instances in this category. Notably, although datasets like HumanEval and MBPP are not specialized for performance evaluation, they play an important role in ensuring the correctness of the optimized code~\cite{DBLP:journals/corr/abs-2405-15189, DBLP:conf/nips/ShinnCGNY23, DBLP:journals/corr/abs-2406-12502, li2024falcon, zhang2024codedpo}. Other datasets, such as PIE, EffiBench, and CodeContests, often emphasize runtime and memory of different languages, using large-scale coding repositories like CodeNet, Leetcode, and Aizu online judge~\cite{gao2024searchbasedllmscodeoptimization, DBLP:conf/nips/MadaanTGHGW0DPY23, huang2024effi, DBLP:conf/apsec/ShirafujiOSMW23}. While competitive programming datasets offer high accessibility and suitability for benchmarking optimization methods in controlled environments, they may not represent the complexity of real-world programs, potentially limiting the generalizability of the findings.

\subsubsection{Genereal SE datasets}
The second largest category, with 13 instances, were datasets that are designed for various software engineering tasks, derived from sources like GitHub, StackOverflow, or industry case studies. Examples include a dataset curated by~\citet{DBLP:conf/sigsoft/GargMCSW22}, which features real-world performance improvement changes made by C\# developers to open-source repositories on GitHub; an LLVM-IR dataset compiled from 610k handwritten C/C++ from open source projects~\cite{DBLP:conf/sigsoft/GargMCSW22}; and AnghaBench, a large dataset with one million C-language samples and meta data on their runtimes and code sizes~\cite{DBLP:conf/hpec/GuoM22}. The usage of general SE datasets highlights the effort to validate optimization methods in practical scenarios. 

\subsubsection{Compiler datasets}
Compiler datasets were used in six compiler-related tasks, typically involving input, output code and optimization sequences. As examples, PolyBench, which consists of 30 synthetic computational kernels for various tasks such as linear algebra, matrix operations and physics simulations, were used to evaluate the performance of compiler optimizations in three studies~\cite{DBLP:journals/corr/abs-2408-03408, DBLP:journals/corr/abs-2404-18864, DBLP:journals/pacmpl/YeZSS23}, and an LLVM-IR dataset with 1M functions were used to train and evaluate a LLaMA-2-7B model to search for the optimal compiler optimization passes \cite{DBLP:journals/corr/abs-2309-07062, DBLP:journals/corr/abs-2403-14714}.

\subsubsection{Data science datasets}
Lastly, two studies evaluated their models using code in data science problems. In particular, DS-1000 is a dataset for evaluating the execution performance of LLM-generated code, with a thousand test cases from data science (DS) problems \cite{zhang2024codedpo}, and Big-DS-1000 extends upon it by increasing the data size of test cases by 10 to 1000 times, allowing for more rigorous assessments of code optimization methods \cite{qu2024dynamic}.

\keybox{
\faIcon{search} \textit{\textbf{Finding \thefindingcount:} Various datasets were employed, reflecting the diverse focuses on domains, languages and performance metrics in code optimization studies, where competitive coding datasets stood out as the most common category with 35 instances, yet it may not capture the complexity of real-world programs and limit the generalizability of the findings.}
\addtocounter{findingcount}{1}
}

\suggestbox{
\faIcon{thumbs-up} \textit{\textbf{Recommendation \thesuggestioncount:} Incorporating a diverse range of datasets can help comprehensively assess the strengths and limitations of different optimization techniques under varying conditions.}
\addtocounter{suggestioncount}{1}
}

\subsection{RQ4.2: Were they Evaluated Using Real-World Data?}
\label{subsec:real_world_evaluation}
To further explore the existing evaluation methods, we investigate how many studies evaluate their methods with real-world data in Figure~\ref{fig:real_world}, since this may help researchers and practitioners recognize the significance of real-world evaluations. 

In particular, a significant portion of the studies (36, or 68\%) \textbf{were not evaluated on sophisticated real-life software projects}, but only competitive programming code~\cite{DBLP:conf/nips/MadaanTGHGW0DPY23, DBLP:journals/corr/abs-2401-08500, DBLP:journals/corr/abs-2312-05657, DBLP:journals/corr/abs-2407-14044, DBLP:journals/corr/abs-2405-15189}, synthetic programs~\cite{DBLP:journals/corr/abs-2406-04693, DBLP:journals/corr/abs-2404-18864, DBLP:journals/pacmpl/YeZSS23, DBLP:journals/corr/abs-2408-03408}, or optimization algorithms~\cite{van2024loop, hemberg2024evolving, DBLP:journals/nature/RomeraParedesBNBKDREWFKF24, van2024llamea}. These results reflect an obvious preference for non-real-world datasets like competitive coding or synthetic ones, possibly due to their rich availability and reproducibility and an overlook of real-world validations, which are critical for demonstrating the robustness and applicability of optimization methods in more complex scenarios.

Among the remaining studies, 12 (23\%) incorporated only \textbf{code segments from real-world scenarios} such as software repositories~\cite{DBLP:conf/hpec/GuoM22, peng2024large, li2024instructcoderinstructiontuninglarge, shivashankar2024betterpythonprogrammingall}, compiler optimization problems~\cite{DBLP:conf/euromlsys/GrubisicSSLMC24, DBLP:journals/corr/abs-2309-07062, DBLP:journals/corr/abs-2403-14714}, and data science tasks~\cite{qu2024dynamic, zhang2024codedpo}, yet they still do not fully capture the complexity of real-world software projects or codebases, leaving potential threats to the validity of the evaluation results.

Furthermore, only 9\% of studies evaluated the proposed code optimization approaches using \textbf{full real-world projects}~\cite{choi2024iterative, yao2024rtlrewritermethodologieslargemodels, DBLP:journals/corr/abs-2306-17077, DBLP:journals/concurrency/HanLDLZ24, DBLP:conf/sigsoft/GargMCSW22}. For instance,~\citet{choi2024iterative} utilized Defects4J, a dataset of open-source Java projects with metadata on code issues, complexity and test cases, designed to advance software engineering research; \citet{DBLP:conf/sigsoft/GargMCSW22} collected 45k C\# repositories on GitHub with performance-improving commits, aiming to estimate the impact of code optimizations on various performance metrics like CPU and memory allocation; and \citet{DBLP:journals/concurrency/HanLDLZ24} leveraged three comprehensive case studies in industrial settings to demonstrate that LMs can translate structured text to C with enhanced readability and maintainability that meet industrial standards. 

This limitation in using comprehensive real-world projects illustrates a notable gap in the literature, revealing challenges in obtaining such datasets, due to the dynamic and noisy nature of realistic environments~\cite{gong2024predicting}. Hence, it is crucial to perform thorough validations to ensure the accuracy, reliability, and relevance of the data~\cite{DBLP:journals/corr/abs-2407-14044}.

\keybox{
\faIcon{search} \textit{\textbf{Finding \thefindingcount:} The majority of studies (68\%) did not evaluate code optimization methods with real-world programs, while only 9\% studies employed full real-world projects, with the rest focusing on real-world code snippets, highlighting a notable gap in the literature.}
\addtocounter{findingcount}{1}
}

\suggestbox{
\faIcon{thumbs-up} \textit{\textbf{Recommendation \thesuggestioncount:} Future studies should prioritize the integration of real-world datasets in the evaluation of code optimization techniques, particularly full-scale projects, to ensure their practical relevance and robustness, while being mindful of data quality and reliability.}
\addtocounter{suggestioncount}{1}
}

\begin{figure}[!t]
\centering
\begin{minipage}{0.35\textwidth}
\begin{tikzpicture}[scale=0.8]
\tikzstyle{every node}=[font=\large]
\definecolor{beaublue}{rgb}{0.74, 0.83, 0.9}
\definecolor{applegreen}{rgb}{0.55, 0.71, 0.0}
\definecolor{dollarbill}{rgb}{0.52, 0.73, 0.4}
\definecolor{ao(english)}{rgb}{0.0, 0.5, 0.0}
\definecolor{amaranth}{rgb}{0.9, 0.17, 0.31}
\definecolor{cerublue}{rgb}{0.16, 0.32, 0.75}
\definecolor{frenchblue}{rgb}{0.0, 0.45, 0.73}
\definecolor{iceberg}{rgb}{0.44, 0.65, 0.82}
  
  
    \pie[
        /tikz/every pin/.style={align=center},
        text=pin,
        rotate=15,
        explode=0, 
        color={cerublue!70, iceberg!50, frenchblue!50},
        ]
        {
        68/\texttt{No} (36),
        23/\texttt{Snippets} (12),
        9/\texttt{Projects} (5)
        }

    \end{tikzpicture}
\vspace{-0.3cm}
  \caption{Distribution of evaluation using real-world code.}
 \label{fig:real_world}
  \end{minipage}
~\hspace{0.5cm}
\begin{minipage}{0.4\textwidth}
\begin{tikzpicture}
\begin{axis}[
    xbar=0.05cm,
    width=10cm,
    height=6cm,
    xmax=26,
    bar width=0.4cm,
    bar shift=0cm, 
 enlarge y limits=0.1,
 enlarge x limits=0.02,
    legend style={at={(0.8,0.36)},anchor=north,legend columns=1,font=\small},
    legend cell align={left},
    legend image code/.code={
    \draw [#1] (0cm,-0.1cm) rectangle (0.15cm,0.15cm); },
   xlabel={\# studies},
     x label style={at={(0.5,0.02)},font=\Large},
     nodes near coords,
          point meta=explicit symbolic,
          every node near coord/.append style={anchor=west,font=\large},
        symbolic y coords={G,F,E,D,C,B,A},
         y tick label style={font=\large},
          x tick label style={font=\large},
          yticklabels={,,,,},
         ytick=data
    ]

\addplot[fill=cerublue!70, draw=black, forget plot] coordinates {(22,A)[22 (\%PI)]};
\addplot[fill=cerublue!70, draw=black, forget plot] coordinates {(17,B)[17 (PI)]};
\addplot[fill=cerublue!70, draw=black] coordinates {(12,C)[12 (Speedup)]};
\addlegendentry{Performance gain} 
\addplot[fill=iceberg!70, draw=black] coordinates {(10,D)[10 (\%OPT)]};
\addplot[fill=iceberg!70, draw=black, forget plot] coordinates {(2,E) [2 (AOCC)]};
\addlegendentry{Task-specific} 
\addplot[fill=red!60, draw=black] coordinates {(1,F) [1 (IOCCB)]};
\addplot[fill=red!60, draw=black] coordinates {(1,G) [1 (NPI)]};
\addlegendentry{Self-proposed} 

\end{axis}
\end{tikzpicture}
  \caption{Distribution of evaluation metrics (one study might be in multiple categories).}
 \label{fig:evaluation_metric}
  \end{minipage}
 \end{figure}

\subsection{RQ4.3: What Metrics Were Used for Evaluation?}
\label{subsec:evaluation_metrics}
Evaluation metrics standardize the assessment of optimization techniques, ensuring that comparisons are meaningful and reliable. Thus, we examine the evaluation metrics in this subsection and classify them into three categories as illustrated in Figure~\ref{fig:evaluation_metric}, aiming to aid readers in selecting the most appropriate and effective optimization techniques for their specific needs.

\subsubsection{Performance gain metrics}
A total of 51 evaluation metrics were performance gain-related, within which 22 studies specifically utilized \textbf{percentage performance improvement (\%PI)}, which is determined by the difference between the original and optimized performance normalized by the original performance, offering easy comparisons across different test cases and performance metrics~\cite{DBLP:journals/corr/abs-2405-15189, DBLP:conf/nips/ShinnCGNY23, DBLP:conf/apsec/ShirafujiOSMW23, DBLP:conf/euromlsys/GrubisicSSLMC24, DBLP:journals/nature/RomeraParedesBNBKDREWFKF24}. Another 17 studies used \textbf{performance improvement (PI)}, calculated by the absolute difference in performance metrics before and after optimization, serving as a straightforward view of performance gained~\cite{hemberg2024evolving, han2024generating, jain2022jigsaw, palkowski2024gpt}. Additionally, 12 studies employed the \textbf{speedup (SP)} metric, which is computed by dividing the original performance by the optimized performance, indicating the increase factor in performance~\cite{gao2024searchbasedllmscodeoptimization, DBLP:journals/corr/abs-2404-18864, DBLP:journals/corr/abs-2406-04693, DBLP:journals/corr/abs-2312-05657}. Overall, these metrics are useful for broad comparisons, but can oversimplify the complex characteristics of performance.

\subsubsection{Task-specific metrics}
Recognizing the diversity of code optimization objectives, 12 primary studies leveraged task-specific evaluation metrics. Specifically, 10 of them assessed the effectiveness of LMs by calculating the \textbf{percentage of optimized programs (\%OPT)}, which represents the proportion of code snippets that were successfully optimized by the LM~\cite{DBLP:journals/corr/abs-2309-14846, gao2024searchbasedllmscodeoptimization, DBLP:journals/corr/abs-2109-13498, DBLP:conf/nips/MadaanTGHGW0DPY23, DBLP:journals/corr/abs-2312-05657}. Subsequently, two studies employed \textbf{Area Over the Convergence Curve (AOCC)} to evaluate how quickly and effectively optimization algorithms converge to optimal or near-optimal solutions~\cite{van2024llamea, van2024loop}. Compared to general performance gain metrics, these metrics may provide tailored insights into task-specific optimization challenges.

\subsubsection{Self-proposed metrics}
Additionally,~\citet{pan2024measuringcodeefficiencyoptimization} designed two customized metrics for evaluations for specific research goals. Particularly, \textbf{Isomorphic Optimal Comparison CodeBLEU (IOCCB)} assesses the similarity between the LM-generated optimized code and an ideally optimized version, serving as a measure of the LM's ability to achieve optimal or near-optimal solutions, and \textbf{Normalized Performance Index (NPI)} evaluates the performance of code in terms of its execution time relative to the maximum and minimum execution times of codes that achieve the same functionality, reflecting the relative efficiency of the code.

\keybox{
\faIcon{search} \textit{\textbf{Finding \thefindingcount:} Performance gain metrics like \%PI, PI, and SP offer broad comparisons of optimization effectiveness (51 instances); task-specific metrics, including \%OPT and AOCC, provide focused insights tailored to specific tasks (12 instances); and custom metrics assess optimization capabilities based on unique research needs (two studies).}
\addtocounter{findingcount}{1}
}

\suggestbox{
\faIcon{thumbs-up} \textit{\textbf{Recommendation \thesuggestioncount:} (1) It's essential to combine different types of metrics to obtain comprehensive evaluations of code optimization techniques. (2) Researchers could develop and adopt new metrics that better capture the multifaceted nature of optimization challenges and solutions.}
\addtocounter{suggestioncount}{1}
}

\section{Challenges and Future Directions}
\label{sec:challenges_future_directions}
\newcounter{ChallengeCount}
\setcounter{ChallengeCount}{1}

Despite the rapid advancements of LM-based code optimization in recent years, our survey results reveal several key knowledge gaps that persist. In this section, we will outline these critical open challenges and propose promising future directions to address them.

\subsection{Challenge \theChallengeCount\addtocounter{ChallengeCount}{1}: Balancing Model Complexity and Practicality}
As discussed in Section~\ref{subsec:base_LMs}, the sizes of LMs have been steadily increasing---the most popular GPT-4 models proposed in 2024 have approximately 1.8T parameters. This trend towards larger and more complex LMs requires substantial computational resources for generating and optimizing code, posing a significant challenge for their application in code optimization. 

Meanwhile, as modern software systems grow in complexity and size, it becomes critical for LM-based code optimization methods to handle large-scale codebases in real-world scenarios. Therefore, a notable challenge remains in balancing the complexity and capabilities of LMs with the necessity for practical and cost-effective solutions.

\subsubsection{Future directions}
\textbf{Model compression}: 
Recent research has shown that model compression can significantly reduce model sizes without significant performance loss~\cite{gholami2024can}. A survey by~\citet{zhu2023survey} covered recent model compression methods such as pruning redundant parameters, quantizing weights, or using knowledge distillation to train a smaller model (student) to replicate the behavior of a larger model (teacher), aiming to bridge the gap of balancing model complexity with practicality. For instance, \citet{sun2024simple} introduced Wanda, a method that prunes weights based on their magnitudes and corresponding input activations, enhancing efficiency while maintaining competitive results. However, compression might remove critical parameters that are crucial for understanding complex code semantics or performing precise optimizations~\cite{zhu2023survey}. Therefore, it is vital for future studies to investigate how compression affects LM behavior and its application to code optimization.

\textbf{Ensembling smaller LMs}: Ensembling techniques combine multiple smaller language models to collectively achieve the performance of a single large model while offering greater modularity and flexibility. For example, \citet{chen2024role} provided a comprehensive review of the advantages, challenges, and practical applications of small models (SMs), highlighting how SMs can be combined in ensemble frameworks to approximate the performance of larger models while maintaining modularity and efficiency. Similarly, \citet{lu2024merge} presented a survey on recent methods in this domain, such as LM merging, ensembling, and cooperation, showcasing their advancements in overcoming individual model limitations and achieving higher efficiency, adaptability, and performance. Nonetheless, the complexity of managing and deploying ensembles may offset the computational savings gained from using smaller models. Thus, ensuring effective communication and knowledge sharing between models is essential for future studies to avoid inefficiencies or redundant computations.

\subsection{Challenge \theChallengeCount\addtocounter{ChallengeCount}{1}: Limited Interaction with External Systems}
As highlighted in Section~\ref{subsec:common_challenges}, most LM-based code optimization methods operate in isolated computational environments, unlike human programmers who can dynamically search the Internet, utilize external code analysis tools, and consult with other experts to produce optimal code modifications. Although several studies have employed techniques like contextual prompting, feedback-based, and simple agentic approaches to address this issue as shown in Section~\ref{subsec:code_optimization_methods}, their interactions with external systems remain highly limited and lack scalability, as they fail to integrate seamlessly with external environments and tools, such as expert knowledge, predictive models, and IDEs \cite{DBLP:conf/nips/ShinnCGNY23,   wei2024improving, peng2024large}, thereby resulting in suboptimal optimizations. Hence, it is crucial for future studies to enhance the interaction capabilities of LMs, enabling them to function more effectively like human programmers in real-world software development scenarios.

\subsubsection{Future directions} \textbf{Agentic LMs}: LM-based agents extend the capabilities of standalone LMs by incorporating features that allow them to dynamically perceive and utilize external resources and tools, engage in multi-agent systems and human interaction, thereby tackling complex tasks like code optimization more effectively. As supported by a recent survey on LM agents \cite{liu2024large}, these agents can perform complex, end-to-end software engineering tasks that normal LMs may struggle with, and they can work together, leveraging specialized resources to improve the efficiency and effectiveness of LMs in several coding tasks. However, integrating multiple components and complex mechanisms can lead to increased computational resource requirements, especially in large-scale applications. Moreover, issues related to robustness, security, and fairness are often underexplored, necessitating future research to address these challenges comprehensively.

\subsection{Challenge \theChallengeCount\addtocounter{ChallengeCount}{1}: Limited Generalizability Across Languages and Performance Metrics}
For code optimization techniques to be broadly applicable, they must generalize well across different programming languages and performance metrics. However, variations in syntax, semantics, and performance characteristics can hinder the transferability of optimization strategies~\cite{DBLP:conf/iclr/ShypulaMZ0GYHNR24, DBLP:journals/corr/abs-2406-11935}. Thus, there is a gap in applying learned optimizations effectively across different programming languages and performance metrics. As proven by Sections~\ref{subsec:programming_languages} and \ref{subsec:performance_metrics}, 81\% and 79\% of the primary studies focused on optimizing one single language and performance metric, respectively.

\subsubsection{Future directions} \textbf{Cross-lingual models tailored for code optimization}: Existing studies have developed models trained on multi-lingual datasets such as PolyCoder to improve their ability to generalize across languages~\cite{DBLP:conf/pldi/polycoder22}. However, these models are primarily designed for general code generation tasks and lack the specialization needed for code performance optimization. Future research could focus on adapting these models to learn optimization patterns that are effective across multiple languages, addressing syntactic and semantic variations. 

\textbf{Multi-objective code optimization}: Multi-objective optimization frameworks like NSGA-II have been successfully applied for evolutionary and optimization algorithms~\cite{wu2024evolutionary}, yet, they are rarely applied in code optimization contexts, as the interaction between multiple performance metrics---such as runtime efficiency, memory usage, and energy consumption---is often conflicting and difficult to balance~\cite{peng2024large}. Hence, challenges for future research lie in understanding the trade-offs between these performance metrics and enhancing LM-based optimization methods to achieve a well-balanced optimization.

\subsection{Challenge \theChallengeCount\addtocounter{ChallengeCount}{1}: Limited Evaluation on Real-World Code}
According to our survey results in Section~\ref{subsec:real_world_evaluation}, only 32\% of primary studies tested their code optimization methods on real-world data, which suggests there could be a gap between LMs’ theoretical optimization capabilities and their practical applicability to real-world codebases. Real-world codebases, which often contain complex, legacy, or poorly documented code, are often far more complex than competitive programming and synthetic datasets, leading to degradations of LM-based code optimization approaches~\cite{DBLP:journals/corr/abs-2109-13498,   choi2024iterative}. Consequently, bridging this gap is crucial for future studies to adopt LMs in real-world code optimization and software development scenarios.

\subsubsection{Future directions} \textbf{Establishing standardized real-world benchmark}: One of the critical future directions is to establish standardized, publicly available benchmarks tailored to real-world codebases. As we show in Table~\ref{tb:datasets}, even though 10 datasets have focused on general SE code optimization, four of them are not open-source, and the rest either focus on a single language or single domain, limiting the applicability. Therefore, future efforts should reflect the diverse and complex nature of industrial code, including legacy systems and poorly documented environments. Such benchmarks should incorporate metrics that evaluate optimization outcomes comprehensively, such as scalability, compatibility, efficiency, and maintainability under practical constraints. 


\textbf{Enabling context-aware optimization}: Context-aware optimization involves leveraging multimodal inputs, such as documentation, code comments, and version history, to tailor solutions effectively~\cite{pan2024ecodemasteringefficientcode}, or employing agentic approaches to enable LMs to dynamically interact with the environments to iteratively refine their understanding of the code~\cite{DBLP:journals/corr/abs-2406-04693,   sun2024autosat}. However, integrating these complex modules with existing LM architectures may pose a major technical difficulty.








\subsection{Challenge \theChallengeCount\addtocounter{ChallengeCount}{1}: Trust and Reliability in AI-Driven Code Optimization}
As illustrated by~\citet{yao2024rtlrewritermethodologieslargemodels} and~\citet{sun2024autosat}, LMs inherently exhibit random, inconsistent, and hallucinate answers, which may reduce the trustworthiness and reliability of the optimized code in real-world software systems, hence, human expertise is still essential to validate, interpret, and refine these recommendations. Indeed,~\citet{M2024humanAI} have demonstrated that the integration of human oversight and AI capabilities fosters a productive synergy, wherein humans bring domain knowledge and critical judgment while AI offers computational efficiency and predictive insights. Ultimately, there is a need for effective collaboration between human developers and LMs to achieve optimal code optimization outcomes.

\subsubsection{Future directions} \textbf{Reinforcement learning from human feedback (RLHF)}: Existing code optimization methods have leveraged human expertise through directed preference optimization, as shown in Table~\ref{tb:code_optimization}, which aligns the model's outputs with human preferences via fine-tuning. Extending this approach, RLHF frameworks can utilize human feedback as a dynamic reward signal to guide LMs for specific optimization tasks~\cite{wang2024RLHF}. However, human-provided feedback may introduce inconsistencies or cultural biases that affect the fairness and neutrality of the model, which should be considered by future studies.



\section{Conclusion}
\label{sec:conclusion}

We have presented a systematic literature review on the application of language models (LMs) in code optimization, synthesizing data from over 50 recently published, high-quality and relevant studies. While it is impossible to provide a definitive cataloger of all research, we have tried to provide a comprehensive and accessible survey of the main research areas and future directions. Specifically, we identify five key knowledge gaps that may hinder the field's development, including the challenge of balancing model complexity with practical applicability, and the pressing need for greater generalizability and trust in AI-driven code optimization. 
Addressing these gaps requires further research on more effective techniques and the establishment of standardized evaluation benchmarks. By mapping the evolving landscape of LMs in code optimization, this survey provides a roadmap to overcome current limitations and accelerate advancements in AI-driven software development. LMs and deep learning are not panaceas for all challenges in software engineering and code optimization. LMs must learn from the data they are provided, which inherently shapes their capabilities and limitations. Contrary to concerns that these technologies might reduce the role of software engineers, they instead present new opportunities for enhanced creativity and the exploration of new research frontiers.


\bibliographystyle{ACM-Reference-Format}
\bibliography{references}

\end{document}